\documentclass[twocolumn]{aastex701}
\usepackage{amssymb}
\usepackage{pifont}
\newcommand{\cmark}{\checkmark}
\newcommand{\xmark}{\ding{55}}
\usepackage{amsmath}
\usepackage{printlen}
\usepackage{pgf}
\usepackage[figuresright]{rotating}
\usepackage{booktabs}
\usepackage{pifont}

\submitjournal{ApJ}

\begin{document}

\title{%
   The \chandra-\gaia\ Catalog of Counterparts: \\
    Resolving ambiguous {\gaia} matches to X-ray sources in the {\chandra} Source Catalog using Machine Learning}

\correspondingauthor{Vinay L.\ Kashyap}
\email{vkashyap@cfa.harvard.edu}

\author[0009-0000-5483-2652]{V. Samuel Pérez-Díaz}
\affiliation{Center for Astrophysics $\mid$ Harvard \& Smithsonian, 60 Garden St,
Cambridge MA 02138, USA}
\affiliation{Harvard John A. Paulson School of Engineering and Applied Sciences, 150 Western Ave, Allston, MA 02134, USA}
\affiliation{Universidad del Rosario, School of Engineering, Science and Technology, Cll. 12C No. 6-25,
Bogotá, Colombia}
\affiliation{The NSF AI Institute for Artificial Intelligence and Fundamental Interactions,
Cambridge MA 02139, USA}
\affiliation{New York University, Courant Institute, 60 5th Avenue, New York NY, USA}
\email{v.perez.diaz@nyu.edu}

\author[0000-0002-3869-7996]{Vinay L.\ Kashyap}
\affiliation{Center for Astrophysics $\mid$ Harvard \& Smithsonian, 60 Garden St,
Cambridge MA 02138, USA}
\email{vkashyap@cfa.harvard.edu}

\author[0000-0002-8751-9010]{Joshua D.\ Ingram}
\affiliation{Center for Astrophysics $\mid$ Harvard \& Smithsonian, 60 Garden St,
Cambridge MA 02138, USA}
\affiliation{Carnegie Mellon University, 5000 Forbes Avenue, Pittsburgh, PA 15213}
\affiliation{New College of Florida, 5800 Bayshore Road, Sarasota, FL 34243, USA}
\email{joshingram2018@gmail.com}

\author[0000-0001-5028-5161]{David Fouhey}
\affiliation{New York University, Courant Institute, 60 5th Avenue, New York NY, USA}
\email{david.fouhey@nyu.edu}

\author[0000-0002-5069-0324]{Juan Rafael Martínez-Galarza}
\affiliation{Center for Astrophysics $\mid$ Harvard \& Smithsonian, 60 Garden St, Cambridge MA 02138, USA}
\email{jmartine@cfa.harvard.edu}

\author[0000-0002-8178-8463]{Pavlos Protopapas}
\affiliation{Harvard John A. Paulson School of Engineering and Applied Sciences, 150 Western Ave, Allston, MA 02134, USA}
\email{pavlos@seas.harvard.edu}

\author[0000-0002-0210-2276]{Jeremy J.\ Drake}
\affiliation{Lockheed Martin Solar and Astrophysics Laboratory, 3251 Hanover St, Palo Alto, CA 94304, USA}
\email{jeremy.1.drake@lmco.com}

\author[0000-0003-0560-8392]{Dong-Woo Kim}
\affiliation{Center for Astrophysics $\mid$ Harvard \& Smithsonian, 60 Garden St, Cambridge MA 02138, USA}
\email{dkim@cfa.harvard.edu}

\author[0000-0002-8791-6286]{Cecilia Garraffo}
\affiliation{Center for Astrophysics $\mid$ Harvard \& Smithsonian, 60 Garden St, Cambridge MA 02138, USA}
\email{cgarraffo@cfa.harvard.edu}

% New commands
\newcommand{\chandra}{{\sl Chandra}}
\newcommand{\gaia}{{\sl Gaia}}
\newcommand{\pNWAY}{p_{\text{i}}}                     % nway probability p_i
\newcommand{\pANY}{p_{\text{any}}}                  % any-match probability
\newcommand{\pML}{p_{\text{ML}}}                    % ML probability
\newcommand{\offaxis}{\theta}                       % off-axis angle
\newcommand{\sep}{\Delta r}                         % angular separation
\newcommand{\nmatch}{n_{\mathrm{match}}}                   % number of matches
\newcommand{\nway}{\texttt{NWAY}}

% Photometry and astrometry variables
\newcommand{\gmag}{g_\textrm{mag}}                              % G magnitude  
\newcommand{\bpmag}{B_{P}}                        % BP magnitude
\newcommand{\rpmag}{R_{P}}                        % RP magnitude
\newcommand{\bprp}{B_{P}-R_{P}}                  % BP-RP color
\newcommand{\bpg}{B_{P}-g_\textrm{mag}}                        % BP-G color
\newcommand{\grp}{g_\textrm{mag}-R_{P}}                        % G-RP color
\newcommand{\pmra}{\mu_{\alpha*}}                  % proper motion in RA
\newcommand{\pmdec}{\mu_{\delta}}                  % proper motion in Dec

% Machine learning specific terms
\newcommand{\pmatchindvar}{\texttt{p\_match\_ind}}  % ML match probability variable name
\newcommand{\pany}{p_{\text{any}}}                 % any-match probability (simple form)
\newcommand{\nwaypany}{\texttt{p\_any}}            % NWAY p_any variable name
\newcommand{\nwaypi}{\texttt{p\_i}}                % NWAY p_i variable name
\newcommand{\lgbm}{\texttt{LightGBM}}              % LightGBM classifier
\newcommand{\ml}{\textit{machine learning}}        % machine learning

% Catalog names
\newcommand{\chandracat}{{\chandra\ Source Catalog}}  % Chandra Source Catalog
\newcommand{\gaiadr}{\gaia\ Data Release}       % Gaia Data Release
\newcommand{\cscversion}{\texttt{CSC}\,v2.1}                    % CSC version 2.1
\newcommand{\gdrversion}{\texttt{GDR}3}                      % Gaia DR3
\newcommand{\rmax}{{r_\mathrm{max}}}

\newcommand{\vlk}[1]{\textcolor{red}{[VLK] {#1}}}
\newcommand{\vspd}[1]{\textcolor{blue}{[VSPD] {#1}}}
\newcommand{\jdi}[1]{\textcolor{blue}{[JDI] {#1}}}
\newcommand{\cg}[1]{\textcolor{red}{[CG] {#1}}}

\newcommand{\changed}[1]{{\textcolor{red}{\textbf{#1}}}}

\begin{abstract}
{We present a framework to cross-match sources from the Chandra Source Catalog (CSC v2.1) with optical sources from Gaia Data Release 3.} {Unlike purely spatial approaches, we use source properties such as magnitudes, colors, and distances to identify true counterparts, detect chance coincidences, and resolve ambiguities when multiple plausible candidates exist. We define a training set of high-confidence matches using \nway{}, a Bayesian cross-matching framework that accounts for positional errors and source densities.} We train a gradient-boosted classifier (\texttt{LightGBM}) on a variety of features from both catalogs. Of the {${\approx}{254}$k} unique X-ray sources, we find counterparts for {${\approx}{113}$k} sources, of which plausible multiple counterparts are found for {${\approx}{7}$k}. We find no counterparts for {${\approx}{20}$k} sources for which separation-based cross-matching does find a match, and attribute half of these to chance coincidences. {We validate the pipeline on the Chandra Orion Ultradeep Project (COUP), where the machine-learning matches reproduce 95\% of NWAY cross-matches without using any positional information. We release a catalog of the ${\approx}113$k Chandra-Gaia counterparts, together with ${\approx}7$k alternative matches and ${\approx}20$k ambiguous NWAY associations, supporting future population studies of sources detectable by both Chandra and Gaia. We discuss limitations and provide a generalization of the framework that is applicable in other cross-matching scenarios.}
\end{abstract}


\section{Introduction} \label{sec:intro}

The correct identification of a source in a catalog, allowing for evaluations of statistical completeness \citep{2024ApJS..275...30T}, is a critical factor in astronomical inference.  {Cross-matching} across multiple catalogs is an essential part of this identification process, and is a fundamental driver for accurate classifications and analyses.
With surveys now cataloging billions of sources across multiple wavelengths, the need for reliable and automatic cross-matching systems becomes imperative. 
Traditional approaches to catalog cross-matching predominantly rely on positional information; the simplest strategy is nearest-neighbor matching within a fixed radius. 
Although effective for optical and infrared surveys with precise astrometry, these methods struggle in densely populated sky regions or when positional uncertainties are large, or when there is an imbalance in astrometric precision between the catalogs being matched.
Robust Bayesian cross-matching frameworks have been developed by treating matching as an inference problem: \citet{budavariprobabilisticcrossidentificationastronomical2008} formalized the computation of match probabilities using Bayes theorem, incorporating positional uncertainties and priors on source properties. 
This approach has been extended in several works and is the foundation for more advanced Bayesian methods in cross-matching \citep{czeslalikelihoodskybayesian2023}. 
More recently, methods leveraging multi-wavelength properties have emerged, significantly improving cross-match reliability. 
A leading example is \nway{} \citep{salvatofindingcounterpartsallsky2018}, which extends the Bayesian framework by incorporating observational properties like optical and infrared magnitudes as {\sl priors}. 
\nway{} provides posterior probabilities for every potential counterpart, including the possibility of no counterpart at all. 
\nway{} has become a key baseline tool in recent astronomy cross-matching efforts, notably adopted in studies such as for the eROSITA mission \citep{brunnererositafinalequatorialdepth2022, schneidererositafinalequatorialdepth2022, kimchandrasourcecatalog2023}. 
Comprehensive reviews on cross-matching methodologies can be found in \cite{salvatofindingcounterpartsallsky2018, budavariprobabilisticrecordlinkage2015, czeslalikelihoodskybayesian2023}.

While Bayesian multi-parametric cross-matching improves significantly on purely positional methods, it relies heavily on carefully constructed priors for each included physical property. 
Furthermore, adjustments to computed probabilities are done on a population level, and not on an individual basis.  
Building these priors can be challenging and labor-intensive, particularly given the complexity and diversity of astronomical datasets. 
With catalogs becoming larger and more precise, there is now a clear need for a systematic and flexible approach that fully leverages multidimensional catalog data. 
Recent efforts have addressed this challenge by integrating \ml{} (ML) techniques directly into the cross-matching pipeline. The eROSITA team developed several applications of this concept, by using a \ml{} classifier and feeding it as prior to \nway{} \citep{salvatoerositafinalequatorialdepth2022, collaborationeuclidquickdata2025}, and to provide final classifications of stellar sources \citet{schneidererositafinalequatorialdepth2022}. 
\cite{freundstellarcontentrosat2022, freundsrgerositaallsky2024a} extended these approaches to identify stellar (coronal) X-ray emitters across the entire ROSAT all-sky survey and the eROSITA all-sky survey (eRASS1), respectively. 
Cross-matching X-ray sources with optical catalogs will help with filling in the gaps on the evolution of stellar magnetic activity by allowing confrontations with state-of-the-art models. 

\chandra's long mission duration has allowed for the development of an extensive database of targeted detections of X-ray sources covering a large portion of the sky \citep[CSC; \chandra\ Source Catalog,][]{evanschandrasourcecatalog2020,evanschandrasourcecatalog2024}{.} %, known as the Chandra Source Catalog (CSC). 
In the most recent release, CSC2.1 \citep{evanschandrasourcecatalog2024}, 1.3 million individual detections (comprising 407,806 unique X-ray sources) are 
available with photometric, variability, and spectroscopic information. %released with spectroscopy, photometry, variability, and spectral fits. 
Several groups have cross-matched the CSC with other catalogs like SDSS \citep{greenchandrasourcecatalog2024}, ALLWISE and \gaia\
\citep{rotschandrasourcecatalog2018, rotsupdatecrossmatchingcsc2020}.
Of these, the \gaia\ \citep{gaiacollaborationgaiamission2016,marresegaiadatarelease2019} Data Release 3 %e.g., the Gaia Data Release 3 
\citep[\gaia\ DR3][]{vallenarigaiadatarelease2023}, comprising almost 1.8~billion sources,
is of special interest because of its comprehensive depth, sky coverage, high positional precision, availability of reliable parallaxes, and focus on galactic (stellar) populations.  
Matching CSC to \gaia\ DR3 allows subsequent follow-up studies of stellar evolution to be carried out reliably. 

In this work, we describe a method to assign reliable \gaia\ counterparts to X-ray sources from the \chandracat{}.  The method assigns matches using the {\sl properties} of the individual sources, and we present the results in a comprehensive catalog.  {Since \gaia\ sources are heavily weighted towards being stars, such a catalog supports important studies} in stellar astrophysics, such as constructing detailed luminosity functions and investigating the physical processes driving the evolution of stellar X-ray activity. The cross-match procedure we present is easily generalizable and can be extended to include other combinations of catalogs, as well as other astrophysical objects detectable by both \chandra\ and \gaia, such as quasars, active galactic nuclei (AGN), and other extragalactic or variable sources \citep{rimoldinigaiadatarelease2023}. Specifically, we describe:
\begin{itemize}
    \item A cross-match procedure based initially on positional information and uncertainties, using the \nway{} algorithm \citep{salvatofindingcounterpartsallsky2018}.
    \item A statistical analysis of \gaia\ magnitude distributions for counterparts to X-ray sources, providing insights into limitations of purely positional cross-matching and the construction of a reliable dataset for \ml{}.
    \item  A machine-learning based pipeline for identifying counterparts using catalog properties only. 
    \item An approach to selecting appropriate thresholds for the machine-learning classifier, based on physical and instrumental limitations.
\end{itemize}

We describe the databases we use in Section~\ref{sec:data}, specifically the \chandracat{} v2.1 (Section~\ref{sec:CSC}) and the \gaia~DR\,3 (Section~\ref{sec:GDR3}).  
We describe the steps in our cross-matching process in Section~\ref{sec:cross-matching}, where we {show} how separation-based methods like \nway{} are used (Sections~\ref{sec:NWAY},\ref{sec:limits_to_NWAY}), define our positive and negative sets for training, validation, testing, and verification (Sections~\ref{sec:pos_set}-\ref{sec:verifyset}), describe the properties used for the cross-matching (Section~\ref{sec:features}), and provide details of the ML implementation (Section~\ref{method_lightgbm},\ref{sec:counterpart_selection}).  
We discuss the performance of the algorithm in Section~\ref{sec:performance}, highlighting various measures of validity (Section~\ref{sec:MLperformance}), and compare our results with previously studied surveys (Section~\ref{sec:verification}). 
We present the catalog of cross-matches in Section~\ref{sec:catalog}. 
We discuss its structure in Section~\ref{sec:discuss}, and in particular outline the framework of the process that can be generalized to other scenarios in Section~\ref{sec:framework}.  We summarize this work in Section~\ref{sec:summary}.

\section{Data}\label{sec:data}

\subsection{CSC: The \chandra\ Source Catalog}\label{sec:CSC}

The \chandracat{} (CSC) provides the final collection of summarized properties and data products of the sources detected by the \chandra\ X-ray Observatory in its history. 
In its version 2.1 (\cscversion{}), the catalog covers a sky area of ${\sim}730~\text{deg}^2$ ($\approx$1\% of the sky), and lists $407\,806$ X-ray detections in $15\,533$ observations \citep{evanschandrasourcecatalog2020, evanschandrasourcecatalog2024}. 
Properties are usually measured in $5$ energy bands within the energy range of $0.2-7.0$~keV for the Advanced CCD Imaging Spectrometer (CCD), and one broad energy band ($0.1-10$~keV) for the High Resolution Camera (HRC). 

The catalog lists source positions, detection significance, spatial extent, source intensities, hardness ratios, fit parameters for some nominal spectral models, and temporal variability estimates. 
Due to the unprecedented spatial resolution of \chandra, the \cscversion{} contains the most accurate and precise positions of X-ray sources available, and is therefore the ideal source for multi-wavelength source matching and characterization.

The CSC source and detection properties are organized in three principal tables: Master Sources, Stacked Observation Detections, and Individual Observation Detections. 
In this work, we use the Master Sources table, as it presents the best estimate properties for each distinct X-ray source on the sky, including position. 
This is done by combining the data of all the source's uniquely-associated individual detections and observations. We extract the off-axis angle $\offaxis$ as the minimum average off-axis (property \texttt{theta\_mean}) from all uniquely associated stack detections. 
This represents the most on-axis detection for every given X-ray source, which have a greater weight over the properties in the Master Table. 
In this paper, we refer to the \chandracat{} 2.1 as \cscversion{} or CSC.

\subsection{\gaia: The \gaia\ Data Release}\label{sec:GDR3}

The \gaia~DR\,3 (\gdrversion) \citep{vallenarigaiadatarelease2023} provides accurate astrometry and photometry for ${\sim} 1.8$ billion sources across the full-sky. 
Remarkably, \gaia\ DR3 includes classifications for ${\sim}1.5$ billion sources, many machine-learning based \citep{rimoldinigaiadatarelease2023}, along with parallaxes, proper-motions, and $\bprp$ colors. 
Additionally, mean radial velocities are included for ${\sim}33$ million stars. 
Some $6.6$ million quasar candidates and $4.8$ million galaxy candidates can be encountered in the catalog, with redshift estimates for most quasars and ${\sim}1$ million galaxies. 
\gaia\ DR3 includes sub-milliarcsec precise measurements of positions, proper motions, parallaxes, and with magnitudes in multiple photometric bands, making it ideal for characterizing a diverse set of astrophysical objects. 
In this paper, we refer to this dataset as the \gaia\ DR3, GDR3 or \gaia.

\section{Cross-Matching}
\label{sec:cross-matching}

\subsection{\nway{} as precursor}\label{sec:NWAY}

As a first step of our pipeline, we perform a positional-based cross-match of the complete \cscversion{} and \gdrversion{} using \nway{}\footnote{{We use NWAY v4.5.2; see \url{https://github.com/JohannesBuchner/nway}.  There are also other equivalent methods, such as XMATCH \citep{rotsupdatecrossmatchingcsc2020, Rots2025xmatch}, but for the sake of simplicity and ease of comparisons we limit our analysis to NWAY.}} \citep{salvatofindingcounterpartsallsky2018}. 
\nway{} provides the probability of a counterpart being the correct one (referred to here as $\pNWAY$ or \nwaypi{}) and the probability that any of the possible counterparts is the real one ($\pANY$ or \nwaypany{}). 
For a description of these probabilities, see Appendix \ref{app:nway}. 

\begin{figure*}
    \centering
    \includegraphics[width=\linewidth]{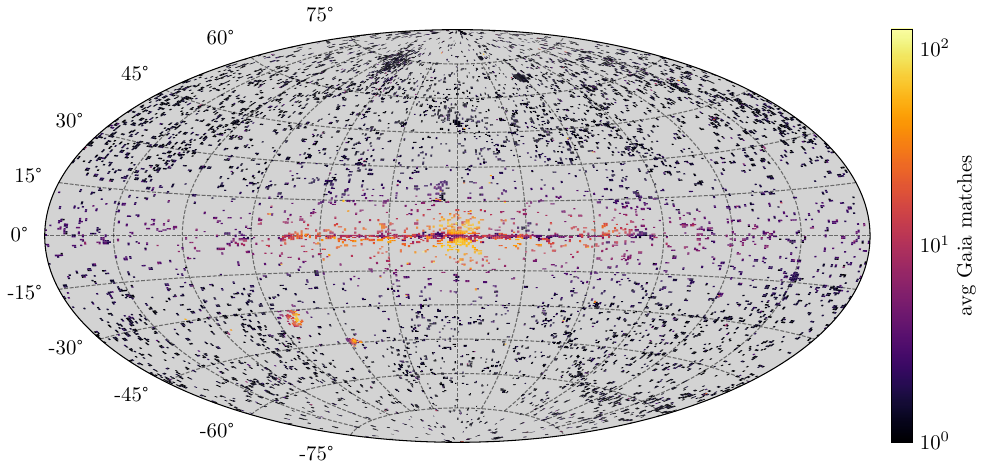}
    \caption{Sky histogram-distribution of all CSC2.1 sources with a potential \gaia\ counterpart within $15''$. Each bin is colored based on the average number of potential \gaia\ counterparts per X-ray source (avg $\nmatch$). The color scale is logarithmic.}
    \label{fig:sky-hist}
\end{figure*}

\begin{figure}
    \centering
    \includegraphics[width=\columnwidth]{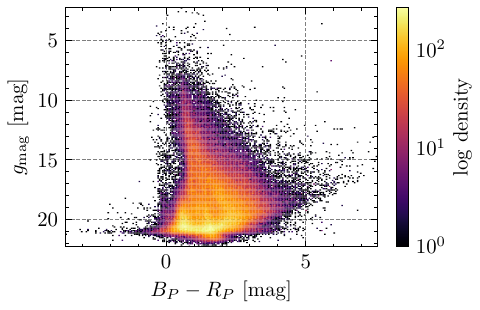}
    \caption{{Color-magnitude diagram for all \gaia\ sources which were selected as best matches by \nway{} (i.e., $\pANY\ge{0.5}$ and $\max\{\pNWAY\}$). %(i.e., $\texttt{match\_label} = 1$).
    }
    The surface densities of ($\gmag,\bprp$) are shown as a hexbin plot (constructed using {\tt matplotlib.pyplot.hexbin}) on a logarithmic color scale (see color bar at right).}
    \label{fig:bprpvsgmag}
\end{figure}

For each X-ray source, we search for optical counterparts within a radius of $15\arcsec$, which covers several times the maximum positional error of \chandra\ (see Appendix~\ref{app:thresholds}).  {This set of \gaia{} sources forms the initial candidate pool to test for matches.}  To account for the non-uniform sky coverage of \chandra\ compared to \gaia\ (former is pointing based-detection, the latter is an all-sky survey), we set the prior completeness factor $c$ (i.e., the prior expected fraction of reliable counterparts) to $0.5$. 
As a result, we get a preliminary cross-match based only on positional information, accounting for sky coverage and positional errors in both missions. We keep every possible association of the output for later analysis.  A census of the associations is in Table~\ref{tab:csc_gaia_census}.  
{Of the $407\,806$ unique X-ray sources in CSCv2.1, $254\,309$ have Gaia sources within 15$''$, yielding a pool of $2\,402\,340$ candidates.  The remainder have no Gaia candidates within the search radius and are not considered further in this work.}
On average, each unique \chandra\ source {in the output table} has ${\sim}9$ potential counterparts. 
About one third of the sample has a single possible counterpart, and about two thirds have less than 5. 
These numbers provide a foundation to understand the complexity of the particular matching problem at hand.
In Figure \ref{fig:sky-hist} we present the sky distribution of all the CSC2.1 sources from this output, with the density of potential optical candidates per X-ray source as the color scale. 
We observe some expected trends in this density, such as a dense cluster in the galactic center. 
These dense regions also correspond to where most of the ambiguity is expected. 
Figure \ref{fig:bprpvsgmag} shows the $\gmag$ magnitude vs $\bprp$ for the \gaia\ sources that \nway{} considered as matches to X-ray sources. 
{The densest region of this distribution corresponds to the faintest, redder ($\bprp \geq 0$) sources. This is consistent with stellar X-ray populations, 
where X-ray flux is proportionally higher compared to bolometric flux
% in which later-type stars show stronger coronal X-ray activity 
\citep{freundstellarcontentrosat2022}.}

% Figures

\begin{table}
\centering
\caption{Census of CSC2.1-GDR3 associations after a preliminary \nway{} run.}
\label{tab:csc_gaia_census}
\footnotesize
\begin{tabular}{lc}
\hline\hline\\[-6pt]
Metric  &  Number\\
\hline\\[-6pt]
CSC sources {with potential Gaia candidates} & 254\,309\\
\gaia\ candidates with separation $\leq{15}$\arcsec          & 2\,402\,340\\
CSC sources with $\pANY \geq 0.5$            & 122\,192\\
\gaia\ counterparts for $\pANY \geq 0.5$ & 1\,169\,082\\
CSC sources with $>1$ Gaia candidate within $15''$ & {166,269} \\
Maximum candidates for a single CSC source & 297\\[4pt]
\hline
\end{tabular}
\end{table}

\subsection{Distinguishability of Property Distributions}\label{sec:limits_to_NWAY}

After the initial separation-based cross-match using \nway{}, we investigate whether there are significant distributional differences in the properties among the counterparts for a given X-ray source.
This provides an indication as to when solely separation-based cross-matching becomes unreliable at a distributional level, and acts as a guide for constructing a reliable data set for training the machine learning model for property-based cross-matching.

For the measurement of mean $\gmag$ magnitude, we conduct a two-sided two-sample Kolmogorov-Smirnov (K-S) test \citep{smirnovTableEstimatingGoodness1948} to determine the distinguishability of the property distributions of the most-likely counterparts (highest $\pNWAY$) and least-likely counterparts (lowest $\pNWAY$) at varying separation bins and off-axis angle ranges.
These separation bins are determined by count-based range cutoffs, which are determined by the range required for a bin to contain a sample of $1000$ counterparts for the most-likely counterparts.
We then divide the bins by the off-axis angle ranges of $0\arcmin - 3\arcmin$, $3\arcmin - 5\arcmin$, and $>5\arcmin$, and any sources which do not have separation or off-axis measurements are dropped from the analysis.
After applying these filters to construct various samples, we conduct the two-sided two-sample K-S test for all possible combinations of separation and off-axis angles.
If a $p$-value is less than $0.10$, we consider this to be an indication that the distributions of the most-likely and least-likely counterparts from \nway{} are distinguishable, and thus different, from one another.  In contrast, if $p>0.1$, we consider the two distributions are not distinguishable.

\begin{figure}[t]
\includegraphics[width=\columnwidth]{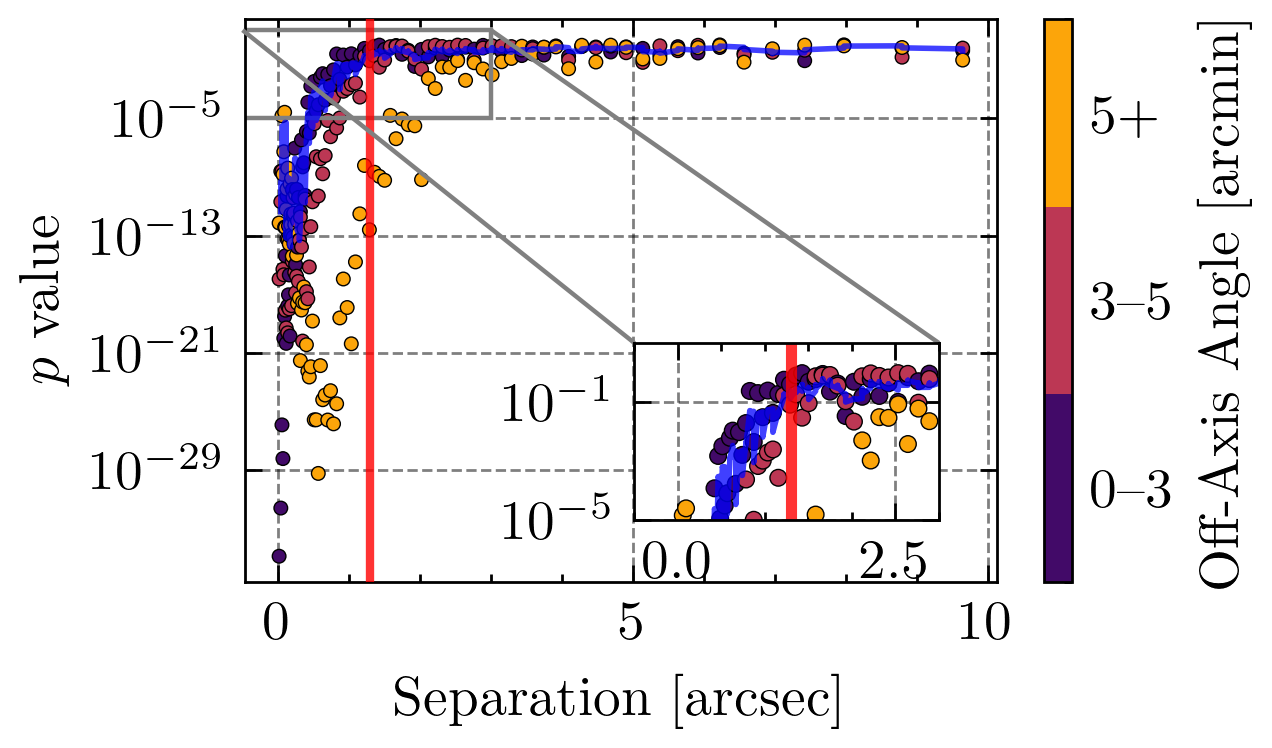}
\centering
\caption{Separation plotted against the $p$-value from the two-sample K-S test to determine after which separation the distributions of mean $\gmag$ magnitude for the most-likely and least-likely counterparts become distinguishable, which is approximately $1.3\arcsec$ or more. The colors of the points correspond to different off-axis angles, the blue solid line is the moving average of the $p$-values across the different off-axis ranges, and the vertical red solid line is the minimum separation for which the (average) $p$-value is at least $0.10$ ($1.3\arcsec$).}
    \label{fig:gmean_distinguish}
\end{figure}

We show the run of $p$-values in Figure~\ref{fig:gmean_distinguish} for these distributions for \gaia\ $\gmag$~band magnitudes.  There is a clear trend in the $p$-values, with the distributions becoming indistinguishable at separations $\gtrsim{1.3}$\arcsec.  
There are small differences when the samples are stratified with off-axis angle $\offaxis$.
This suggests that separation-based \nway{} matches become unreliable when the counterparts are more than {${\approx}{1.3}\arcsec$} apart.  We thus construct our positive set (see below) by considering only those counterparts that are closer than {this critical separation} threshold.

\subsection{Cross-matching as binary classification}\label{sec:crossmatch_classify}

As we demonstrate above, matches based solely on spatial separation become unreliable on average beyond a certain critical separation.  To resolve the ambiguities that result from that, we formulate the task of catalog cross-matching as a binary classification problem that takes advantage of additional catalog properties.  For each candidate association between a source in the primary catalog (e.g., CSC) and an entry in the secondary catalog (e.g., \gaia), we define a binary target variable $y \in \{0,1\}$, where $y = 1$ if the candidate is a true counterpart, and $y = 0$ otherwise.

The model takes as input a set of mixed properties from \textit{both} catalogs (e.g., magnitudes, flux ratios, color indices, classification labels, etc), and outputs a real-valued score $\pML$.  We treat the score $\pML$ as a measure of the degree of certainty that the candidate association is correct.  The final predicted class $\hat{y}$ is then obtained by applying a threshold to $\pML$, and further selection criteria over properties.  We train a machine learning model to follow this scheme, without using any positional information so that it learns a signal based only on properties.  In order to do that, we define a series of positive and negative sets (see below).

In a broader sense, recent work treats the binary classification problem as modeling the question ``\textit{Is this source an X-ray emitter?}" \citep{salvatoerositafinalequatorialdepth2022, collaborationeuclidquickdata2025} for feeding it into \nway{} as a prior. 
In this work, we formulate the question ``\textit{How likely is it that this set of properties are associated with the same source?}". 
We treat this step as a \textit{separate} matching criterion (property-based) to \nway{} (separation-based). 
Thus, we do not feed the score as a prior but rather use the score as a complementary discrimination factor after the spatial cross-match.

\subsection{The Positive Set}
\label{sec:pos_set}

To construct the positive set, we select the most probable \gaia\ counterpart for each CSC2.1 source. Specifically, for each X-ray source, we take the candidate with the maximum $\pNWAY$, and apply the following selection criteria:
\begin{eqnarray}
    &&\texttt{separation} \leq 1.3\arcsec \nonumber \\
    &&\mathrm{for~off~axis~angle} ~\theta\in[0,3]~\mathrm{arcmin} \nonumber \\
    &&\pANY\geq{0.5} ~\textrm{and}~ \pNWAY\geq{0.9} \,.
\end{eqnarray}
The value of $1.3\arcsec$ for the separation threshold was derived from the analysis described in Section~\ref{sec:limits_to_NWAY}.
{Choosing $\pANY\geq{0.5}$ and $\pNWAY>0.9$ ensures that the sample is dominated by high-confidence, unambiguous associations.}
Each pair that satisfies these constraints is considered a true match and is assigned the label $y = 1$.

\subsection{The Negative Set}\label{sec:neg_set}

Negative samples are defined as candidate \gaia\ associations to the same CSC2.1 sources that do not meet the positive selection criteria. We define three types of negatives:

\paragraph{Clear negatives} For each CSC source in the positive set, we identify the \textit{non-selected} candidate with the largest separation, requiring at least $\texttt{separation} \geq 5''$. These cases are labeled as \texttt{clear\_negative}.

\paragraph{Intermediate negatives} We also include all other non-selected candidates for the same sources that satisfy $\texttt{separation} \geq 5''$, but are not the farthest. These cases are labeled as \texttt{intermediate}.

\paragraph{Random negatives} We augment the clear and intermediate cases with randomly sampled \gaia\ sources. Specifically, for each positive match, we sample $k$ unrelated \gaia\ sources uniformly from a pre-filtered sky catalog that excludes the $15''$ radius around any CSC source footprint center. These samples are {coupled} to the CSC sources by repeating each X-ray row $k$ times and pairing it with $k$ independent \gaia\ entries.  This results in a set of synthetic negative associations that are guaranteed to be spatially unrelated to any known X-ray source. To make their status explicit, we assign placeholder values to properties that depend on real possible associations (e.g., \texttt{separation}). These cases are labeled as \texttt{random}.

Only candidates associated with CSC sources that appear in the positive set for the $[0, 3] \arcmin$ off-axis range are considered. All negatives are labeled with $y = 0$.

Following this procedure, we obtain a clean and consistent training dataset with one reliable positive match and multiple realistic negative candidates per X-ray source. It captures both trivial and challenging examples of non-associations.  {Note that the negative set does not include X-ray sources that are not in the positive set.  This ensures that the classifier learns to distinguish true \gaia\ counterparts from chance associations for sources of the same type, under the same conditions, and this similarity to the positive set enables it to achieve high out-of-distribution performance \citep{Ursu2024.06.17.599333}.}

\begin{figure}
    \centering
    \includegraphics[width=\columnwidth]{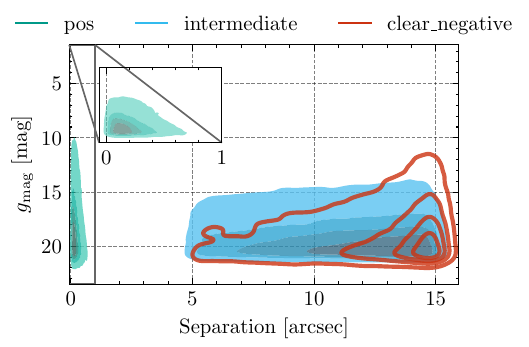}
    \caption{Separation of positive and negative sets.  Densities of the \gaia\ $\gmag$ versus angular separation of putative \gaia\ matches are shown for the positive set (green shaded densities at left and in the inset), the {\sl intermediate negatives} (blue shaded densities at right) and {\sl clear negatives} (red contours superposed on the intermediate negatives) are shown.  The inset of the positive set zooms in to the separation range of 0-1\arcsec along the X-axis but covers the same $\gmag$ range along the Y-axis as the main plot.  Notice that the marginalized $\gmag$ distributions are clearly different for the positive and negative sets.
    }
    \label{fig:kde_gmag_vs_separation}
\end{figure}

In Figure \ref{fig:kde_gmag_vs_separation} we show a density plot of the $\gmag$ magnitude versus angular separation to the X-ray source for all \gaia\ sources in the the positive and negative set. The negative set is subdivided in the \texttt{intermediate} and \texttt{clear\_negative}. We do not include the \texttt{random} cases, as by construction they lack a separation.

\subsection{Training, Validation, and Test Sets}\label{sec:trainvalidtest}

From the previous process, we get a positive set of $30\,279$ unique associations, and a negative set of $310\,020$ associations without including \texttt{random} negatives. 
Already, we observe a very high imbalance between positives and negatives in our dataset, so it is necessary to tune the multiplier hyperparameter for the amount of random \texttt{random} negatives in the training set, as well as sample reweighting to account for the class imbalance. 
We discuss how this is accounted for in Section~\ref{method_lightgbm}.

{We randomly split the positive set into training, validation, and test subsets using an 80/20 approach.  We set aside the first 20\% of the sources as test set for the final evaluation of the model performance.  The remainder is further split in the same proportions, with 80\% of the 80\% (64\% of the total) subset is used for training, and 20\% of the 80\% (16\% of the total) is used for validation.  The validation and test subsets include all \gaia\ candidates within 15$''$ of each CSC source; i.e., we compute the individual NWAY probabilities $\pNWAY$ for each \gaia\ candidate.  In order to facilitate the evaluation of metrics like precision and F1 score (see below), the best \gaia\ counterpart for each X-ray source is labeled $1$, and the remainder are labeled $0$.}

\subsection{Verification set}\label{sec:verifyset}

The data split described before was specifically set up for the machine learning component of the pipeline. This is just a step of the counterpart selection criteria, and thus we need a way of verifying how the full pipeline performs on a carefully selected set. {We use the COUP survey \citep[Chandra Orion Ultradeep Project;][]{getmanchandraorionultradeep2005,2005ApJS..160..353G} as our verification set, and present an analysis of the cross-matches in it in Section~\ref{sec:verification}}.

\subsection{Features}
\label{sec:features}
{Features were selected from properties available in both catalogs and external studies. We included as many as physically motivated, as the gradient boosting framework naturally down-weights uninformative properties via feature importance.}

Table~\ref{tab:features} summarizes the full set of 32 properties we use as input to a model. Alongside the catalogue properties, we include machine-learning based classification labels from \citet{yangclassifyingunidentifiedxray2022, perez-diazunsupervisedmachinelearning2024}. 
These labels might give the model an informed prior on hard cases it might otherwise mislabel. 

\begin{table*}
\centering
\caption{{Features used in the \lgbm\ classifier, and the split importance factor for each feature.} 
%List of features used in the classifier. % The three top features based on \lgbm{} split importance are bolded.
}
\label{tab:features}
\begin{tabular}{llc}
\hline\hline\\[-6pt]
Feature & Description & {Split Importance (\%)}\\
\hline\\[-6pt]
\multicolumn{3}{l}{\textit{Photometry (Gaia)}}\\[2pt]\hline\\[-6pt]

phot\_g\_mean\_mag          & Mean $\gmag$ magnitude                                         & {4.3} \\
phot\_bp\_mean\_mag         & Mean $\bpmag$ magnitude                                        & {2.8} \\
phot\_rp\_mean\_mag         & Mean $\rpmag$ magnitude                                        & {3.3} \\
phot\_g\_mean\_flux         & Integrated $\gmag$-band flux [$\text{electrons s}^{-1}$]       & {3.6} \\
phot\_bp\_mean\_flux        & Integrated $\bpmag$ flux [$\text{electrons s}^{-1}$]           & {2.2} \\
phot\_rp\_mean\_flux        & Integrated $\rpmag$ flux [$\text{electrons s}^{-1}$]           & {3.2} \\[4pt]

\hline\\[-6pt]
\multicolumn{3}{l}{\textit{Derived colors (\gaia)}}\\[2pt]\hline\\[-6pt]
bp\_rp                      & $\bprp$ color                                                  & {4.9} \\
bp\_g                       & $\bpg$ color                                                   & {3.7} \\
{{g\_rp}}    & {{$\grp$ color}}                                & {{5.3}} \\[4pt]

\hline\\[-6pt]
\multicolumn{3}{l}{\textit{X-ray flux and colors based on CSC passbands (\chandra)}}\\[2pt]\hline\\[-6pt]
{{photflux\_aper\_b}} & {{X-ray net photon flux in the broad band [photons~s$^{-1}$~cm$^{-2}$]}} & {{6.9}} \\
hard\_hs                    & Hardness ratio, hard ($H$) and soft ($S$) bands~ $HR_{HS} = \frac{H-S}{H+S}$    & {4.7} \\
hard\_hm                    & Hardness ratio, hard ($H$) and medium ($M$) bands~ $HR_{HM} = \frac{H-M}{H+M}$  & {4.0} \\
hard\_ms                    & Hardness ratio, medium ($M$) and soft ($S$) bands~ $HR_{MS} = \frac{M-S}{M+S}$  & {4.3} \\[4pt]

\hline\\[-6pt]
\multicolumn{3}{l}{\textit{Intra/inter-observation variability (\chandra)}}\\[2pt]\hline\\[-6pt]
var\_intra\_prob\_b         & Intra-obs variability probability, broad band                  & {3.9} \\
var\_intra\_index\_b        & Intra-obs variability index, broad band                        & {0.9} \\
var\_inter\_prob\_b         & Inter-obs variability probability, broad band                  & {2.8} \\
var\_inter\_index\_b        & Inter-obs variability index, broad band                        & {0.7} \\
var\_inter\_sigma\_b        & Inter-obs flux variability, broad band                         & {3.5} \\[4pt]

\hline\\[-6pt]
\multicolumn{3}{l}{\textit{Extended/Flux/Variability Flag (\chandra\ and \gaia)}}\\[2pt]\hline\\[-6pt]
extent\_flag                & X-ray source is extended flag                                  & {0.2} \\
phot\_variable\_flag        & \gaia\ variability flag                                        & {0.4} \\[4pt]

\hline\\[-6pt]
\multicolumn{3}{l}{\textit{Catalogue classifications (External and \gaia)}}\\[2pt]\hline\\[-6pt]
yangetal\_gcs\_class        & Supervised-learning class \citep{yangclassifyingunidentifiedxray2022}         & {3.8} \\
yangetal\_training\_class   & Training classes \citep{yangclassifyingunidentifiedxray2022}                  & {0.5} \\
perezdiazetal\_class        & Unsupervised-learning class \citep{perez-diazunsupervisedmachinelearning2024} & {0.2} \\
classprob\_dsc\_combmod\_quasar & Probability of being a quasar                                 & {4.9} \\
classprob\_dsc\_combmod\_galaxy & Probability of being a galaxy                                 & {4.9} \\
classprob\_dsc\_combmod\_star   & Probability of being a star                                   & {4.4} \\[4pt]

\hline\\[-6pt]
\multicolumn{3}{l}{\textit{Astrometry (Gaia)}}\\[2pt]\hline\\[-6pt]
parallax                    & Parallax [mas]                                                 & {4.2} \\
parallax\_error             & Uncertainty on parallax [mas]                                  & {4.3} \\
{{sqrt(pmra\string^2 + pmdec\string^2)}} & {{Proper motion [mas~yr$^{-1}$]}} & {{5.0}} \\
radial\_velocity            & Radial velocity [km~s$^{-1}$]                                  & {0.4} \\
vbroad                      & Spectral line broadening [km~s$^{-1}$]                         & {0.1} \\
distance\_gspphot\          & Distance estimate [pc]                                         & {1.7} \\[4pt]
\hline
\end{tabular}
\end{table*}

\section{Implementation}\label{sec:implementation}

\subsection{LightGBM}
\label{method_lightgbm}

In order to obtain a cross-matching score based only on non-spatial catalog properties, a model that extends beyond prior models for certain quantities is needed. 
In this work, we use \lgbm{} \citep{kelightgbmhighlyefficient2017} as our classifier. 

\lgbm{} is a gradient boosting framework based on decision-tree methods, used to solve many machine-learning tasks---like classification---in an efficient way. 
It uses histogram-based splitting, making it significantly faster and more memory-efficient than other similar methods. Unlike classical gradient boosting, \lgbm{} grows trees leaf-wise rather than level-wise, which often leads to better accuracy.
We selected \lgbm{} for two main reasons: its speed, which allows us to run extensive hyperparameter searches efficiently, and its consistently strong performance on tabular datasets like ours. 
{Gradient boosting on tabular data is a well established standard, and performs equivalently or better than other methods \citep{kelightgbmhighlyefficient2017, grinsztajnwhytreebasedmodels2022, hollmannaccuratepredictionssmall2025}.}
Additional advantages include native support for missing values and categorical features.

We train a \lgbm{} model to perform binary classification, assigning a label of 1 (positive) to true matches and 0 (negative) to non-matches. 
We augment the negative set by upsampling five {($k=5$)} \texttt{random} \gaia\ sources per \chandra\ source, following the data procedure described before. Given that our dataset contains a high imbalance, training a model directly on this data would cause it to bias toward predicting negatives. 
To counter this, \lgbm{} adjusts the weight of each positive sample proportionally to the class imbalance ($w_{pos} = \frac{N_{\text{neg}}}{N_\text{pos}}$). 
The final negative set for training include \texttt{clear\_negative}, \texttt{intermediate}, and \texttt{random} negatives. 
This setting up-weights the positive (minority) class, which shifts the output {scalars} away from true probabilities. 
We therefore treat these as uncalibrated ‘scores’ and set our ML-match threshold empirically in Section \ref{sec:MLthresh}.

We optimize hyperparameters using a randomized grid search with 200 iterations and 5-fold cross-validation. 
We select the best model from a set of hyperparameter tuning runs, experimenting with different amounts of \texttt{random} negatives and investigating the impact of including \texttt{intermediate} negatives. Further training details are provided in Appendix \ref{app:model_training}.

\paragraph{Feature Selection} 
{We evaluate the importance of each feature by analyzing their contribution to model performance.  We evaluate it using two methods: (i) permutation feature importance, where each feature is randomly permuted and the model performance is checkde to see how much it degraded, and (ii) using \lgbm{}'s model internal importance score, which measures the number of times a descriptor is used as a splitting node across all decision trees.  We list this split importance value in Table~\ref{tab:features} as a guide to the relative value of each feature.  Both methods consistently find that color indices, magnitudes, and fluxes are the most important features.  This is an expected result, given known relations between X-ray and optical flux and the clustering of objects in color-magnitude space.}
We find no performance gain from removing features, and thus follow the standard approach of retaining all features that may carry signal and are scientifically motivated.

\subsection{Criteria for Selecting Counterparts}
\label{sec:counterpart_selection}

\subsubsection{Limit on maximum separation}
\label{sec:separation_thre}

Our method does not rely directly on the separation of sources in the disparate catalogs.  However, it does use information on spatial closeness in two stages: first, to set up the Positive Set (see Section~\ref{sec:pos_set}), and second, to discard matches at large separations.  The former condition is necessary because a ground truth labeled sample does not exist, and as we show in Figure~\ref{fig:gmean_distinguish}, extremely small separations are a useful diagnostic of matching.  The latter condition is necessary to rule out matches that may be deemed meaningless because they are separated by angular distances larger than expected due to statistical or systematic uncertainties in the respective positions.  The model assigns a score $\pML{}$ to each potential X-ray-optical match.  However, this score alone cannot guarantee reliable counterpart identification by itself.  
Indeed, excellent matches by properties could occur at unjustified large separations (as a trivial example, a match found at a separation of $180^o$ must be discarded as a chance match of the properties).
Thus, we set a maximum separation threshold beyond which ML matches are discarded.  This threshold is used as a binary choice; if $\pML>0.466$ for any \gaia\ source with a smaller separation, it is kept as a counterpart, and discarded entirely if it exceeds the threshold separation.

We set the threshold by utilizing an expected error as a function of off-axis position,
\begin{equation}
\sigma_{\rm err}(\theta)
=\sqrt{(0.1'')^2 + (0.5'')^2 
      + \Bigl(\tfrac{\sigma_{\rm PSF}(\theta)}{3}\Bigr)^{2}
      + \sigma_{\rm PSF}(\theta)^{2}}\,, 
\label{eqn:sigerr}
\end{equation}
where $\sigma_{\text{PSF}}:=R^{\text{ECF}}_{90}/2.15$ (see Equation~\ref{eqn:R_ECF}; a description of how {the 90\% enclosed counts fraction radius,} $R^{\text{ECF}}_{90}$, is estimated is given in Appendix~\ref{app:thresholds}).  This expression is constructed as a combination of several sources of error: the first term represents potential proper motion effects; the second term describes the \chandra\ astrometric precision; the third describes the expected statistical precision for typical sources; and the fourth accounts for possible contamination due to unrecognized overlaps.  Notice that the threshold errs on the side of inclusion, i.e., we seek to minimize the chances of missing a true counterpart.

\begin{figure}
    \centering
    \includegraphics[width=\linewidth]{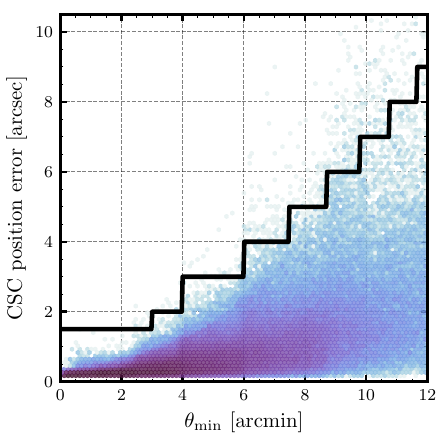}
    \caption{Hexbin density plot of the CSC Positional Error (PE) for all CSC sources with at least one Gaia candidate as a function of the minimum mean off-axis angle.  Here, $\mathrm{PE}=\sqrt{\mathrm{\texttt{errmaj}}^2 + \mathrm{\texttt{errmin}}^2}$, where \texttt{errmaj} and \texttt{errmin} are CSC variables representing the uncertainties along the major and minor axes of the ellipses fitted to the sources.  Also shown as the solid stepped line is our adopted separation threshold $\rmax$, which bounds the CSC PEs.}
    \label{fig:poserrVrmax}
\end{figure}

We then convert $\sigma_{\rm err}(\theta)$ to a discrete-valued, maximum acceptable separation threshold $\rmax$ (in arcseconds) by rounding it up to the next upper integer, i.e., {$\rmax = \lceil\sigma_{\rm err}(\theta)\rceil$}.  We enforce a floor of 1.5\arcsec\ at $\theta < 3\arcmin$ and cap the threshold at 10\arcsec\ for $\theta > 12\arcmin$ to prevent an excessive number of unlikely candidates. {The floor accounts for possible proper motion as well as the documented Chandra absolute astrometric 90\% uncertainty\footnote{See CXC memo at \url{https://cxc.harvard.edu/cal/ASPECT/celmon/}} of 1.2$''$.}
Specifically, the final separation acceptance thresholds are
\begin{eqnarray}
    \rmax &=& 1.5\arcsec~\textrm{for}~\theta\leq3\arcmin \nonumber \\
     &=& 2\arcsec~\textrm{for}~3\arcmin < \theta\leq4\arcmin \nonumber \\
     &=& 3\arcsec~\textrm{for}~4\arcmin < \theta\leq6\arcmin \nonumber \\
     &=& ~~~~ \ldots \nonumber \\
     &=& 10\arcsec~\textrm{for}~\theta\geq12\arcmin \,.
     \label{eqn:rmax}
\end{eqnarray}
Only matches with separations below the computed $\rmax(\theta)$, {\sl and} with ML classification scores above the selected threshold (Section~\ref{sec:MLthresh}), are considered a \ml{} match.  {The separation threshold closely bounds the expected X-ray position errors in the CSC sources we use (see Figure~\ref{fig:poserrVrmax}; position errors dominate other sources of uncertainty at large off-axis positions).  All \gaia\ sources at separations $<\rmax(\theta)$ are tested for cross-matching with the ML model, and are reported in the catalog below (see Section~\ref{sec:catalog}).}

\subsubsection{Machine learning score threshold}\label{sec:MLthresh}

In the previous section, we describe how we select a set of separation thresholds for different off-axis intervals based on \chandra's instrumental limitations. 
This selection is independent of any model score, but determinant on bounding the amount of candidates to a physically-motivated regime. 
This way, the model's selection is fine-tuned. 
We treat it separately from \nway{}'s spatial criteria, as we consider both pipelines independently. 
At the end, we select reliable matches combining the results of both pipelines: spatial, and property/instrument-based. 

For selecting the model's candidates, we must select an appropriate threshold of the model's score in which a candidate is determined to be a match or non-match.  We consider two ways:

\paragraph{Chance coincidence threshold}
We bound our match‐score threshold by computing an estimated probability of chance coincidence, and then empirically selecting a score that reflects this probability from an empirical cumulative-distribution function. 
The \gaia\ density is
$\rho = \frac{N_\textrm{G}}{A_\textrm{sky}}$, where
$N_{\rm G}=1\,811\,709\,771$ and $A_{\rm sky}=4\pi\,\mathrm{sr}\simeq41253\,\mathrm{deg}^2$, and thus $\rho\approx4.4\times10^4~\deg^2$. 
For a matching radius \(r=1.3\arcsec\) (Section \ref{sec:pos_set}) for each X-ray source, we have a total area of $A_\textrm{src} = \pi \cdot (r/3600)^2~\deg^2 \approx 4.1\times10^{-7}~\deg^2$. 
Then, the chance‐coincidence probability is the product of the area of the X-ray sources times the density of \gaia,
\begin{equation}
    P_\textrm{chance} = \rho \cdot A_\textrm{src} \approx 0.0176 \,.
\end{equation}
We then compute the empirical CDF for $\pML{}$ over the {\sl Validation Set}, and linearly interpolate to find the score corresponding to the chance-coincidence probability percentile.  This is shown in Figure~\ref{fig:ecdf}, and any candidate for which $\pML{} <0.466$ is rejected as a likely false match. {Note that $P_{\rm chance}$ uses the global average \gaia\ source density and is therefore a conservative lower bound. Users interested in denser fields (e.g., the Galactic plane) should adopt a correspondingly higher proportion of chance coincidence, which leads to a higher $\pML$ threshold.  The catalogs we construct (see Section~\ref{sec:catalog} below) list the value of $\pML$, and smaller values can be filtered out as needed, reducing the total number of ML matches.
}

\paragraph{{ROC} threshold}\label{sec:ROCAUC}

As an alternative, {one may select the threshold from the classifier's Receiver Operating Characteristic (ROC) curve by maximizing the Youden index, }{$J = \mathrm{TPR}-\mathrm{FPR}$,} where TPR refers to true-positive rate and FPR refers to false-positive rate, or by fixing an acceptable false‐positive rate directly on the validation set. We analyze the results given by this approach as well as the chance-coincidence.

\subsection{Performance}\label{sec:performance}

We analyze the machine learning model test performance in the proxy binary-classification task out of the box (without any additional criteria), and with the additional separation/off-axis thresholds. 
We select a threshold for the machine\ learning model selection, based on the chance-coincidence probability analysis. 
Then, we analyze and compare the machine learning based cross-matches with \nway{}'s separation-based counterparts. 
We deploy the model for {the verification region} and verify its agreements and differences with \nway{}. 
Finally, we deploy the model in the full counterparts dataset. 

\subsection{Baseline ML model performance}\label{sec:MLperformance}

We train the model as described in Section \ref{method_lightgbm}. The model achieves an AUC-ROC {(Area Under the Curve of the ROC curve)} of $0.894$ on the validation set and $0.898$ on the test set. The model is saved and used for the analysis presented in the following sections.

We select the machine-learning threshold as described in Section \ref{sec:MLthresh}.

\begin{figure}
  \centering
  \includegraphics[width=\columnwidth]{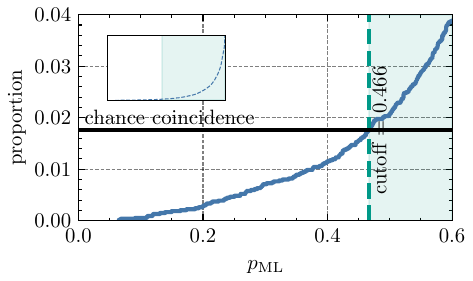}
  \caption{Defining the threshold for acceptance of the ML score for matching, $\pML$.  The empirical cumulative distribution of $\pML$ for the validation set is shown as the blue curve (dashed blue curve in the inset graph).  The proportion that corresponds to chance coincidence is marked as the horizontal black line, and the corresponding $\pML$ score threshold is marked with a vertical green dashed line.  Only cross-matches with scores of $\pML\geq0.466$ are retained as valid counterparts.
  }
  \label{fig:ecdf}
\end{figure}

Figure \ref{fig:ecdf} shows the empirical cumulative distribution function (CDF) of the machine learning probabilities in the a subset of the validation set, and where the computed probability of chance coincidence falls. 
For constructing this CDF, we take a particular subset of the validation set to ensure that the selected counterparts are reliable. 
The criteria are as follows: their \nway{} probabilities $\pNWAY$ and $\pANY$ are greater than $0.9$, and the selected counterpart is the only possible counterpart within the predefined radius ($15\arcsec$); they are isolated counterparts in the sky.

\begin{figure}
  \centering
  \includegraphics[width=\columnwidth]{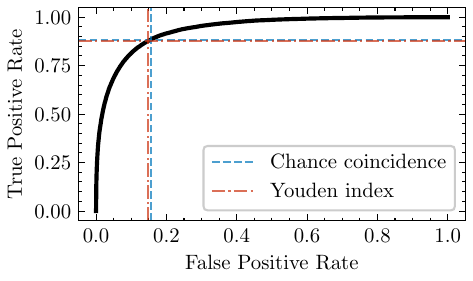}
  \caption{Receiver Operating Characteristic (ROC) curve from the training and validation sets. We highlight the thresholds selected by the chance-coincidence percentile (blue, dashed) and maximizing the Youden Index (red, dashdot).}
  \label{fig:roc_threshold}
\end{figure}

Figure \ref{fig:roc_threshold} shows the ROC curve in the validation set, with the thresholds determined by the chance-coincidence and the Youden index highlighted. 
As observed, the thresholds vary by ${\sim}0.01$, and thus will perform equivalently. 
Therefore, we decide to use the chance-coincidence threshold for the subsequent results in this paper, as it is physically motivated.

\begin{table}
\centering
\caption{Comparing ML and \nway{} matches in the Test Set.
}
\label{tab:nway_ml_comparison}
\begin{tabular}{lr}
\hline\hline
Metric  &  Number\\
\hline
Potential \gaia\ counterparts (within $15''$) & $73\,779$ \\
Size of Test Set & $6\,056$ \\
\nway{} matches & $6\,056$ \\
Recall (\nway{} and ML agree) & $4\,790$ \\
ML and \nway{} matches are equal and unique & $4\,710$ \\
ML does not find a match & $1\,162$ \\
ML finds multiple matches & $80$ \\
ML finds a different match than \nway{} & $0$ \\
\hline
\end{tabular}
\end{table}

With the selected threshold, we derive metrics on the test set. We achieve a precision of $0.31$, and $0.98$ in the negative predictive value. 
We obtain a recall of $0.79$ and an overall F1-score of $0.44$. 
The discrepancy of the precision and recall are a result of the model's lack of spatial properties and the imbalance towards negatives, making it confuse many as positives and biasing the metrics
By incorporating the separation thresholds defined in Section \ref{sec:separation_thre}, we counteract this effect. 

We apply our chosen separation thresholds in combination with the machine-learning threshold to define ML-confident counterparts.
On the test set, this results in a precision of $0.98$, negative predictive value of $0.98$, recall of $0.79$, and F1-score of $0.87$. 
To measure how the ML set overlaps with \nway{}'s matches, we compute the Jaccard index (intersection/union), which results in $ J_{acc} =0.78$. 
The fact that the classifier is not perfectly aligned to \nway{}'s ground truth is a desired effect, as it is expected that a spatial ground-truth has errors by construction. We value, however, that the model is able to recover a signal from the properties. 
Table \ref{tab:nway_ml_comparison} summarizes the per-X-ray-source agreement between the two pipelines.

\subsection{Example: The Orion Nebula Cluster}
\label{sec:verification}

\begin{figure}
    \centering
    \includegraphics[width=\linewidth]{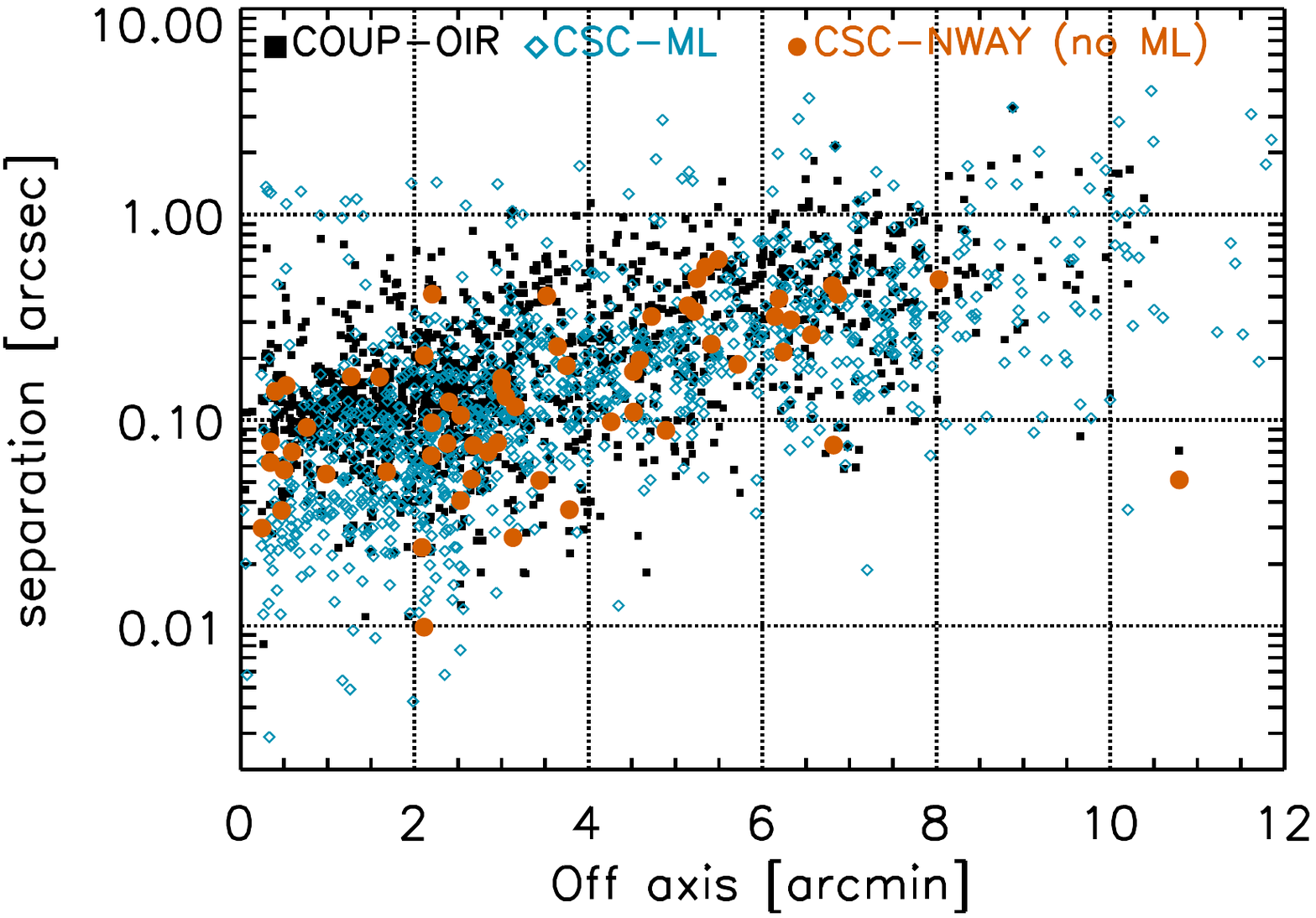}
    \caption{{Comparison of COUP and CSC matches.  The separations between COUP/OIR matches and CSC/Gaia matches are shown as a function of off-axis position.  There is no discernible difference in the scatter of the separations between either COUP/OIR (filled black squares), or the CSC/Gaia matches for NWAY (filled red circles) or ML (open blue diamonds) methods.  They all follow similar trends and similar variances.}}
    \label{fig:COUP2CSC}
\end{figure}

In order to explore the characteristics of our cross-matching method in greater detail, we compare our results to the COUP survey \citep[\chandra\ Orion Ultradeep Project;][]{getmanchandraorionultradeep2005,2005ApJS..160..353G}.  COUP found $1\,616$ X-ray sources in $\approx$840~ks of exposure time, and identified $1\,323$ of them with optical and infra-red (OIR) sources based on proximity.  With CSC, we find $1\,397$ X-ray sources within a radius $11.84'$ of the COUP survey center at (RA,Dec)$_\textrm{2000}$=(05:35:17.0399, -05:23:39.8405), $1\,282$ of which can be confidently associated with COUP X-ray detections\footnote{\label{foot:coupcsc}Even though both COUP and CSC are based on the same \chandra\ observations, there are small algorithmic differences in the data processing, detection algorithm, and merging of results from multiple ObsIDs, which induce systematic shifts in the estimated positions of weak, off-axis, or confused sources.  In order to figure out which CSC sources are also present in COUP, we searched for the nearest COUP source to each CSC source, and found that they were strongly clustered at $0''.12{\pm}0''.14$, but with large outliers, and a dependence on the off-axis position $\theta$.  We fit a straight line to $\log{\textrm{separation}(\theta)}$ minimizing the least absolute deviation for robustness against outliers, adjusted the intercept of the fitted line (but not the slope) to include the peak and a $3$-standard deviations width of the distribution.  The limiting threshold was set at $\textrm{separation}{\le}0.656{\times}10^{0.121\cdot\theta}$~arcsec.}.  

Because COUP relies on spatial proximity to associate X-ray with OIR, and there are significant differences in passband coverage and completeness between the OIR and \gaia\ catalogs, COUP cannot serve as a ground truth survey.  However, as we show in Figure~\ref{fig:COUP2CSC} and Table~\ref{tab:nway_ml_coup}, our method shows a remarkable overlap with COUP, comparable to what would be achieved with \nway{} alone.  For instance, we find that $1,015$ of the CSC sources have the same ML matches as \nway{}, and $917$ also have COUP counterparts with OIR matches.  A total of $978$ CSC sources have matches obtained with both \nway{} and ML techniques.  There are $45$ CSC sources with potential additional counterparts identified with ML, and $60$ \nway{} matches without an ML counterpart.  The similarity of matches is illustrated in Figure~\ref{fig:COUP2CSC}, 
{which shows scatter plots of the separations as a function of off-axis angle for different methods: the original cross-matching of COUP and OIR \citep[black filled squares][]{getmanchandraorionultradeep2005,2005ApJS..160..353G}; the CSC-ML cross-matches (blue open circles); and the CSC-NWAY cross-matches which do not have ML counterparts (red filled circles).  The scatter in all three cases are similar, and have the same trend with off-axis angle, which indicates that they are all sampling from the same distribution of cross-matches and do not exhibit relative biases.}
We emphasize here that since closeness in position is {\sl not} a criterion we use in the ML method to choose a match, a priori there is no reason to expect any of the suggested candidates to achieve an ML match score $\pML>0.466$; yet $\approx{95}$\% of the matches are found to be in agreement.  This is a strong indication that X-ray emitters are recognizable by their optical characteristics.

In addition, we note that $15$ instances are found where ML suggests a different \gaia\ source to be a match than \nway{}.  The detailed cross-matches for both the \nway{} match (including $\pANY$ and $\max{p_i}$) and ML match (including $\max{\pML}$), along with the corresponding \gaia\ $\gmag$ and $\bprp$ are listed in Table~\ref{tab:coup_flips}.  Note that two sources with $\pANY\ll0.5$ are included here because there is a change between $\max{p_i}$ and $\max{\pML}$; these would be excluded from \nway{}-generated catalogs, but included in ML-generated catalogs.  In many cases the optical properties of the two matches are similar, and ML is selecting the \gaia\ match which has better measurements, like the existence of $\bprp$, a smaller parallax error, a $\bprp$ estimate that is less of an outlier, etc.  This suggests that the next \gaia{} data release may resolve these discrepancies by improving the location estimates or eliminating what may be duplicates in the catalog.

\begin{table*}[]
    \centering
    \caption{Summary of the cross-matches over the COUP field}
    \begin{tabular}{lr}
    \hline\hline
    Quantity & Number \\
    \hline
    \gaia\ DR3 sources & 1755 \\
    Total COUP sources & 1616 \\
    COUP sources with OIR matches & 1323 \\
    Total CSC sources & 1397 \\
    CSC sources with -- & \hfil \\
    \quad -- COUP counterparts & 1282 \\
    \quad -- an ML match (as in Table~\ref{tab:best_matches}) & 1015 \\
    \quad -- both \nway{} and ML matches & 978 \\
    \quad -- ML matches that also have COUP OIR IDs & 917 \\
    \quad -- alternate ML matches (as in Table~\ref{tab:alternative_matches}) & 45 \\
    \quad -- ML matches and no \nway{} matches & 75 \\
    \quad -- \nway{} matches but no ML matches (as in Table~\ref{tab:ambiguous_matches}) & 60 \\
    \quad -- different ML matches than \nway{} (Table~\ref{tab:coup_flips}) & 15 \\
    \hline
    \end{tabular}
    \label{tab:nway_ml_coup}
\end{table*}

\begin{table*}[hbt]
    \centering
    \caption{List of flipped cross-matches in the COUP field that have changed between \nway{} and ML.}
    \label{tab:coup_flips}
    \begin{tabular}{lccclccccc}
    \hline\hline
    2CXO ID & $\theta$ & $\pANY$ & Match & Gaia ID & separation & ${p_i}^\dag$ & ${\pML^\dag}$ & $\gmag$ & $\bprp$ \\
    \hfil & [arcmin] & \hfil & algorithm & \hfil & [arcsec] & \hfil & \hfil & [mag] & [mag] \\
    \hline
J053452.7-052753 &  6.12 & $1.5~10^{-8}$ & \nway{} & 3017269470960502656 & 0.905 &     1.000 &    0.714 &  15.19 & NA \\
              \hfil &  \hfil  &  \hfil   &  ML  & 3017269470959732480 & 1.296 & 3.223e-13 &    0.875 &  14.83 & 2.13 \\
J053457.7-052350 &  4.78 &  $3.8~10^{-3}$ & \nway{} & 3017363857172493824 & 0.659 &     1.000 &    0.888 &  14.00 & NA \\
              \hfil &  \hfil  &  \hfil   &  ML  & 3017363857165366400 & 1.862 & 1.100e-44 &    0.929 &  13.39 & 1.98 \\
J053500.1-052301 &  4.18 &     1.00 & \nway{} & 3017364647449498624 & 0.209 &    0.9979 &    0.821 &  12.79 & 1.68 \\
              \hfil &  \hfil  &  \hfil   &  ML  & 3017364647439213184 & 0.497 &  0.002057 &    0.824 &  13.48 & 1.76 \\
J053504.6-052936 &  6.42 &     1.00 & \nway{} & 3017266275504060672 & 0.157 &     1.000 &    0.932 &  15.99 & 2.50 \\
              \hfil &  \hfil  &  \hfil   &  ML  & 3017266275504060800 & 2.920 &     0.000 &    0.946 &  14.39 & 1.81 \\
J053506.1-052212 &  3.00 &     1.00 & \nway{} & 3017364303851923456 & 0.160 &    0.8021 &    0.290 &  19.52 & 0.64 \\
              \hfil &  \hfil  &  \hfil   &  ML  & 3017364299541935744 & 0.255 &    0.1979 &    0.351 &  19.02 & 0.96 \\
J053506.4-053335 &  5.38 &     1.00 & \nway{} & 3017264660593314688 & 0.092 &    0.9068 &    0.539 &  15.82 & 2.03 \\
              \hfil &  \hfil  &  \hfil   &  ML  & 3017264660596368000 & 0.294 &   0.09319 &    0.585 &  15.93 & 1.72 \\
J053514.9-052412 &  0.69 &     1.00 & \nway{} & 3017363960254458112 & 0.110 &     1.000 &    0.786 &  16.09 & 0.86 \\
              \hfil &  \hfil  &  \hfil   &  ML  & 3017363960251436544 & 1.296 & 2.629e-26 &    0.826 &  16.15 & 0.97 \\
J053515.3-052224 &  1.24 &     1.00 & \nway{} & 3017364127743286272 & 0.182 &     1.000 &    0.114 &  17.17 & NA \\
              \hfil &  \hfil  &  \hfil   &  ML  & 3017364132049496832 & 0.606 & 1.697e-05 &    0.674 &  15.96 & 0.96 \\
J053515.4-051943 &  3.90 &     1.00 & \nway{} & 3017366709029832064 & 0.200 &     1.000 &    0.223 &  20.21 & NA \\
              \hfil &  \hfil  &  \hfil   &  ML  & 3017366709022399360 & 1.723 & 3.734e-38 &    0.546 &  19.21 & 0.75 \\
J053515.4-052040 &  2.95 &    0.997 & \nway{} & 3017365918755675648 & 0.400 &     1.000 &    0.326 &  19.33 & NA \\
              \hfil &  \hfil  &  \hfil   &  ML  & 3017365918746317312 & 1.407 & 9.766e-14 &    0.669 &  19.00 & 0.72 \\
J053517.8-051835 &  0.53 &     1.00 & \nway{} & 3017366812104158208 & 0.148 &    0.5720 &    0.189 &  20.83 & 0.34 \\
              \hfil &  \hfil  &  \hfil   &  ML  & 3017366807802891648 & 0.165 &    0.4280 &    0.222 &  21.00 & 1.06 \\
J053524.6-051159 & 11.85 &    0.988 & \nway{} & 3209527325414404864 & 1.214 &    0.6575 &    0.851 &  14.78 & 2.03 \\
              \hfil &  \hfil  &  \hfil   &  ML  & 3209527325412171136 & 2.311 &    0.3425 &    0.956 &  15.86 & 2.96 \\
J053524.6-051909 &  4.85 &    0.997 & \nway{} & 3017366017528606720 & 0.497 &     1.000 &    0.647 &  16.10 & 2.48 \\
              \hfil &  \hfil  &  \hfil   &  ML  & 3017366090553947264 & 2.885 & 8.784e-23 &    0.728 &  18.72 & 1.10 \\
J053540.2-051728 &  5.16 &     1.00 & \nway{} & 3017366365419173632 & 0.226 &     1.000 &    0.108 &  17.90 & NA \\
              \hfil &  \hfil  &  \hfil   &  ML  & 3017366365431829248 & 1.618 & 1.071e-22 &    0.927 &  13.75 & 2.39 \\
J053541.9-052812 &  7.76 &    0.999 & \nway{} & 3017347914245946624 & 0.274 &    0.7642 &    0.889 &  11.75 & 1.33 \\
              \hfil &  \hfil  &  \hfil   &  ML  & 3017347914253414016 & 0.337 &    0.2358 &    0.903 &  11.71 & 1.33 \\

    \hline
    \multicolumn{10}{l}{{$\dag:$ \nway{} probabilities and ML scores for the matching sources.  The best \nway\ matches have $\pNWAY=\max\{\pNWAY\}$ and }} \\
    \multicolumn{10}{l}{{the best ML matches have $\pML=\max\{\pML\}$.}} \\
    \end{tabular}
\end{table*}

\section{The \chandra-\gaia\ Catalog of Counterparts}\label{sec:catalog}

\begin{table*}
    \caption{{Census of CSC2.1-GDR3 cross-matches after ML classification}}\label{tab:csc_gaia_final_products}
    \begin{tabular}{lrl}
    \hline\hline
    Subset & Number & Filter\tablenotemark{a} \\
    \hline
    
    \multicolumn{3}{l}{For CSC sources:--} \\
    -- cross-matched with \nway{} & $122\,192$ & $\pANY\ge{0.5}$ and $\pNWAY=\max\{\pNWAY\}$ \\
    -- matched with both \nway{} and ML & $102\,543$ & $\pANY\ge0.5$ and $\pNWAY=\max\{\pNWAY\}$ and  \\
    \hfil & \hfil & $\pML>{0.466}$ and separation$<\rmax$ \\
    -- with best ML cross-matches (Table~\ref{tab:best_matches}) & $112\,779$ & $\pML>{0.466}$ and separation$<\rmax$ and $\pML=\max\{\pML\}$ \\
    \quad -- exactly 1 \gaia\ counterpart & $106\,603$ & as above and $\#\{<\rmax\}=1$ \\
    \quad -- exactly 2 \gaia\ counterparts & $5\,418$ & as above and $\#\{<\rmax\}=2$ \\
    \quad -- more than 2 \gaia\ counterparts & $775$ & as above and $\#\{<\rmax\}>2$ \\
    -- with matches where both ML and \nway{} agree & $97\,876$ & ($\pML>0.466$ and separation$<\rmax$ and $\pML=\max\{\pML\}$) \\
    \hfil & \hfil & and $(\pANY\ge0.5$ and $\pNWAY=\max\{\pNWAY\}$) and ($\max\{\pML\}=\max\{\pANY\}$) \\
    -- with alternate ML matches (Table~\ref{tab:alternative_matches}) & $6\,176$ & $\#\{$unique($\pML>0.466$ and separation$<\rmax$ and $\pML\ne\max\{\pML\}$)$\}$ \\
    -- with ML match but no \nway{} match & $10\,236$ & $\pML>0.466$ and separation$<\rmax$ and $\pANY<0.5$ \\
    -- with different ML and \nway{} matches & $4\,667$ & $\pML>0.466$ and separation$<\rmax$ and $\max\{\pML\}\ne\max\{\pNWAY\}$ \\
    -- with \nway{} match but no ML match (Table~\ref{tab:ambiguous_matches}) & $19\,649$ & $\pML<0.466$ and ($\pANY>0.5$ and $\pNWAY=\max\{\pNWAY\}$) \\
    
    \multicolumn{3}{l}{For \gaia\ sources:--} \\
    -- all ML matches & $120\,048$ & $\pML>{0.466}$ and separation$<\rmax$ \\
    \quad -- exactly 1 \gaia\ counterpart & $106\,603$ & as above and $\#\{<\rmax\}=1$ \\
    \quad -- exactly 2 \gaia\ counterparts & $10\,836$ & as above and $\#\{<\rmax\}=2$ \\
    \quad -- more than 2 \gaia\ counterparts & $2\,631$ & as above and $\#\{<\rmax\}>2$ \\
    -- with best ML cross-matches (Table~\ref{tab:best_matches}) & $112\,782$ & $\pML>{0.466}$ and separation$<\rmax$ and $\pML=\max\{\pML\}$ \\
    -- alternate ML matches (Table~\ref{tab:alternative_matches}) & $7\,207$ & $\#\{$unique($\pML>0.466$ and separation$<\rmax$ and $\pML\ne\max\{\pML\}$)$\}$ \\
    -- with different ML and \nway{} matches & $6\,923$ & $\pML>0.466$ and separation$<\rmax$ and $\max\{\pML\}\ne\max\{\pNWAY\}$ \\
    -- with \nway{} match but no ML match (Table~\ref{tab:ambiguous_matches}) & $19\,647$ & $\pML<0.466$ and ($\pANY>0.5$ and $\pNWAY=\max\{\pNWAY\}$) \\
    -- with non-unique CSC associations & $208$ & $\#\{$in Table~\ref{tab:best_matches}$\}-\#\{$unique~\gaia-IDs in Table~\ref{tab:best_matches}$\}$ \\
    
    \hline
    \end{tabular}
    \tablenotetext{a}{{The notation $X=\max\{X\}$ means select the element from set $\{X\}$ which is equal to the maximum value in the set (if there are ties, all of them are kept).  Similarly, the notation $X<\max\{X\}$ means all elements with values less than the maximum.  The notation $\#\{\mathrm{condition}\}$ means the number of objects that satisfy the given condition.}}
\end{table*}

We construct our catalog based on the nature of the ML scores between each \chandra-\gaia\ pair.  These are based on three Boolean flags devised to summarize each step in our method:
\begin{itemize}
    \item[(i)] \texttt{flag\_nway\_confident}, which marks the best match by spatial proximity as determined by \nway{}.  It is marked `True' for pairs for which $\pANY\ge{0.5}$, and of all possible counterparts, the one with the highest individual $p_i$.  That is, according to \nway{}, there exists a counterpart, and we select the one that is most likely.
    \item[(ii)] \texttt{flag\_ml\_confident}, which marks as `True' all pairs which exceed the ML score threshold on $\pML$ devised in Section~\ref{sec:MLthresh}.
    \item[(iii)] \texttt{flag\_sep\_ok}, which marks all pairs with a binary indicator that their separation is ${\le}r_{max}(\theta)$ (marked as 1) or $>r_{max}(\theta)$ (marked as 0) (see Equation~\ref{eqn:rmax}).
\end{itemize}
We filter the full catalog, which contains all the \chandra-\gaia\ candidate pairs, based on different combinations of the above flags.  We construct three tables:
\begin{enumerate}
    \item \emph{Best ML matches (Table~\ref{tab:best_matches})} -- lists the best \chandra–\gaia\ pair based on the ML score.  We select CSC sources for which at least one counterpart has both \texttt{flag\_ml\_confident}=True and \texttt{flag\_sep\_ok}=1.  If there are multiple counterparts for a given CSC source, we select the one with the highest ML score $\pML$.
    \item \emph{Alternate ML matches (Table~\ref{tab:alternative_matches})} -- lists all other ML-valid potential \gaia\ counterparts that fall within a plausible separation/off-axis threshold.  These are alternative, valid ML matches to CSC sources already in the best matches table, where we include the entries for which \texttt{flag\_ml\_confident}=True, \texttt{flag\_sep\_ok}=1, and $\pML\ne\max{\pML}$ for any source.
    \item \emph{Ambiguous \nway{} matches (Table~\ref{tab:ambiguous_matches})} -- lists all pairs deemed to be counterparts based on proximity alone, but are not appropriate matches according to ML.  These all have \texttt{flag\_nway\_confident}=True, but with either \texttt{flag\_ml\_confident}=False or \texttt{flag\_sep\_ok}=0.  That is, they have $\pANY\ge{0.5}$ and either $\pML<0.466$ or $\textrm{separation}>\rmax(\theta)$.
\end{enumerate}

\begin{deluxetable*}{ccccccccccccc}
\tablecaption{\textit{CSC2.1--Gaia DR3 Best Matches:}\tablenotemark{a}\\
Illustrative sample of counterparts from the Best Matches catalog. These 10 entries (from 112\,779 total) represent high-confidence matches with the highest ML scores.  The full machine readable table is available on Zenodo \citep{perez_diaz_2026_18652667}.}\label{tab:best_matches}
\tablewidth{0pt}
\tablehead{
\colhead{Name} & \colhead{RA} & \colhead{Dec} & \colhead{$\theta_0$} & \colhead{Gaia ID} & \colhead{$\pNWAY{}$} & \colhead{$\pANY{}$} & \colhead{$\pML{}$} & \colhead{Sep.} & \colhead{$F_N$} & \colhead{$F_M$} & \colhead{$F_S$} & \colhead{$\gmag$} \\
 & \colhead{[deg]} & \colhead{[deg]} & \colhead{[arcmin]} &  &  &  &  & \colhead{[arcsec]} &  &  &  & \colhead{[mag]}
}
\startdata
000000.5+321232 & 0.00 & 32.21 & 2.7 & 2873826084487581440 & $>$0.999 & 0.996 & 0.93 & 0.71 & \cmark & \cmark & \cmark & 20.5 \\
033130.3-205216 & 52.88 & -20.87 & 7.8 & 5100637709724057600 & $>$0.999 & 0.999 & 0.93 & 0.54 & \cmark & \cmark & \cmark & 20.9 \\
063942.0+055414 & 99.93 & 5.90 & 0.98 & 3131126163463999232 & $>$0.999 & 0.997 & 0.66 & 0.21 & \cmark & \cmark & \cmark & 19.3 \\
103517.2-581701 & 158.82 & -58.28 & 5.0 & 5351447410242844288 & $>$0.999 & 0.98 & 0.92 & 0.33 & \cmark & \cmark & \cmark & 16.1 \\
120548.5-085200 & 181.45 & -8.87 & 4.0 & 3582018667782810496 & $>$0.999 & $>$0.999 & 0.99 & 0.20 & \cmark & \cmark & \cmark & 19.9 \\
143030.2-001115 & 217.63 & -0.19 & 4.5 & 3653300766821312256 & $>$0.999 & 0.997 & 0.98 & 0.40 & \cmark & \cmark & \cmark & 19.3 \\
165931.4-400030 & 254.88 & -40.01 & 4.3 & 5967091458400954112 & 0.56 & 0.95 & 0.75 & 0.83 & \cmark & \cmark & \cmark & 17.2 \\
180253.6-230212 & 270.72 & -23.04 & 8.0 & 4069256937710039424 & $>$0.999 & 0.45 & 0.91 & 2.3 & \xmark & \cmark & \cmark & 12.2 \\
201704.3+365058 & 304.27 & 36.85 & 8.2 & 2060562069817318656 & 0.88 & 0.72 & 0.80 & 1.6 & \cmark & \cmark & \cmark & 15.5 \\
235959.8+622403 & 360.00 & 62.40 & 6.0 & 2012805465146791424 & $>$0.999 & 0.95 & 0.91 & 0.40 & \cmark & \cmark & \cmark & 18.9 \\
\enddata
\tablenotetext{a}{Column abbreviations: Name = \texttt{csc21\_name}, RA = \texttt{csc21\_ra}, Dec = \texttt{csc21\_dec}, $\theta_0$ = \texttt{min\_theta\_mean}, Gaia ID = \texttt{gaia3\_source\_id}, $\pNWAY{}$ = \texttt{p\_i}, $\pANY{}$ = \texttt{p\_any}, $\pML{}$ = \texttt{p\_match\_ind}, Sep. = \texttt{separation}, $F_N$ = \texttt{flag\_nway\_confident}, $F_M$ = \texttt{flag\_ml\_confident}, $F_S$ = \texttt{flag\_sep\_ok}, $\gmag$ = \texttt{phot\_g\_mean\_mag}.}
\end{deluxetable*}

The columns in the tables (templates are shown in Table~\ref{tab:best_matches}, and in Appendix~\ref{app:catalog_samples}; for the full tables, see \citealt{perez_diaz_2026_18652667}) are designed to provide the minimum set of parameters needed to refilter the catalog with more conservative settings if desired.  They include the \chandra\ and \gaia\ IDs, the \chandra\ source position, their separations, the \nway{} individual and total probabilities $p_i$ and $\pANY$, the ML scores $\pML$, and the corresponding Boolean flags (in the table templates, True and 1 are represented by a check mark, and False by a cross-out x). 
The full list of columns and their descriptions are presented in Appendix~\ref{app:columns}. 

We present a census of these final products in Table \ref{tab:csc_gaia_final_products}.  Similar to what was seen in the study of the COUP field, there is significant overlap between \nway{} and ML matches.  We find ML matches to $112,779$ CSC sources, of which {$6,176$ have multiple alternative} ML matches.  There are $19,649$ sources which have a probable \nway{} match, but no ML match.

\begin{figure*}
    \centering
    \includegraphics[width=\linewidth]{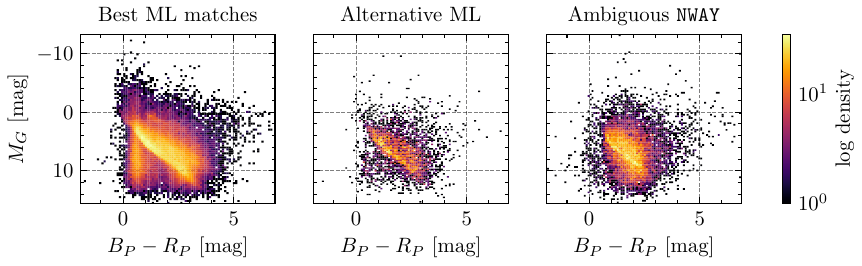}
    \caption{HR diagram for the Gaia sources cross-matched to CSC.  The distance-corrected $\gmag$ is shown as a function of color $\bprp$ for the best ML matches ({\sl left}), for plausible alternate ML matches ({\sl middle}), and for cases with no ML matches but with highly probable \nway{} matches.  In regions of low source concentration along the fringes, sources are marked as individual points; high source concentrations are represented by surface densities.  The morphology of the cross-matches is substantially different for the three cases, though the main sequence is visible in all cases.
    }
    \label{fig:hr_diagram_best_matches_q}
\end{figure*}

\begin{figure*}
    \centering
    \includegraphics[width=\linewidth]{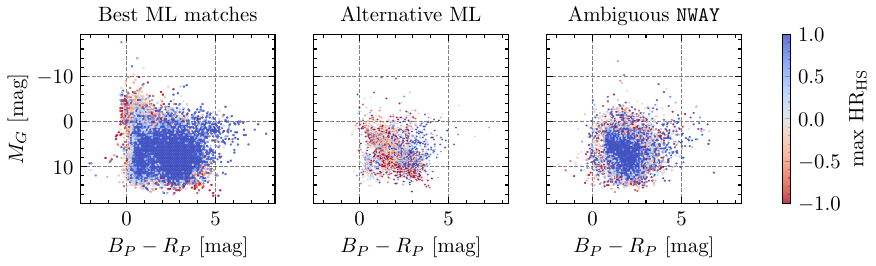}
    \includegraphics[width=\linewidth]{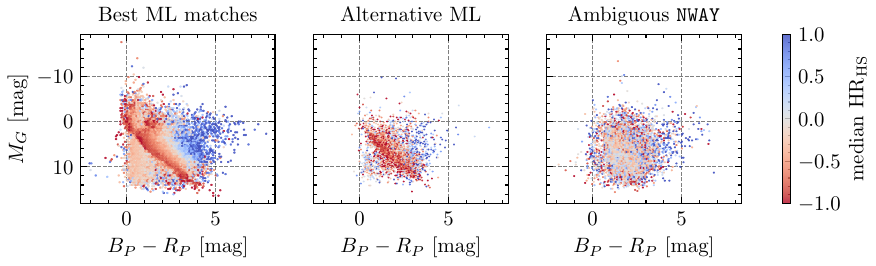}
    \includegraphics[width=\linewidth]{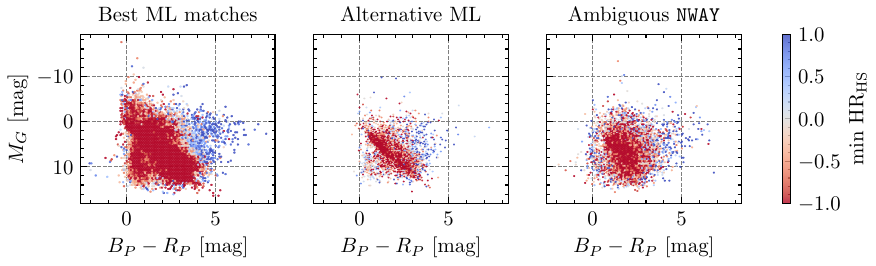}
    \caption{X-ray spectral variations across the color-magnitude space for different types of cross-matches.  Hexbin plots (as in Figure~\ref{fig:bprpvsgmag}) are shown as a function of \gaia\ $\gmag$ and $\bprp$ for the maximum (upper row), median (middle row), and minimum (lower row) of the X-ray $HR_{HS}$ (see Table~\ref{tab:features}).  The counterparts with the best $\pML$ are shown in the left column, those for alternate ML matches are in the middle column, and those with only \nway{} matches are shown in the right column.  The stellar main sequence is clearly visible in the ML matched cases, and the counterparts without reliable ML matches are featureless.
    }
    \label{fig:hr_diagram_test_pos_hardness}
\end{figure*}

\begin{figure*}
    \centering
    \label{fig:separation_vs_p_any_ds}
    \includegraphics[width=\linewidth]{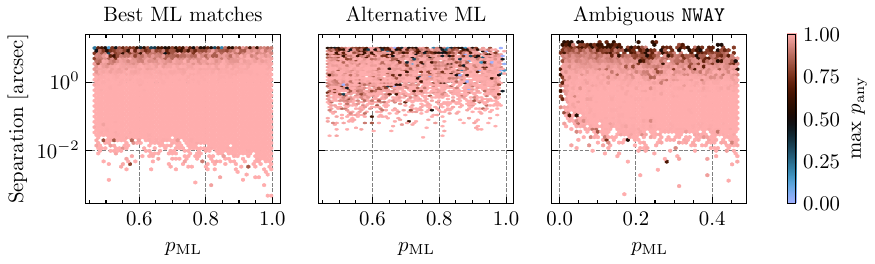}
    \includegraphics[width=\linewidth]{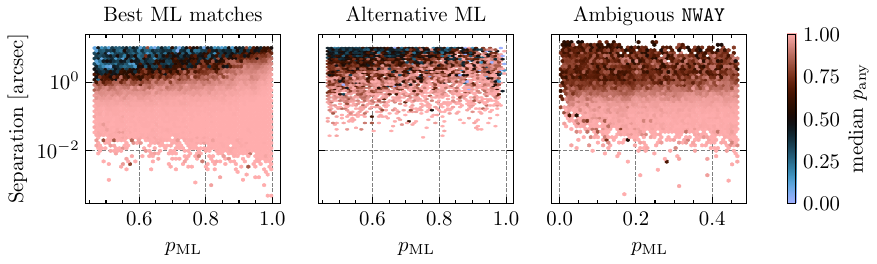}
    \includegraphics[width=\linewidth]{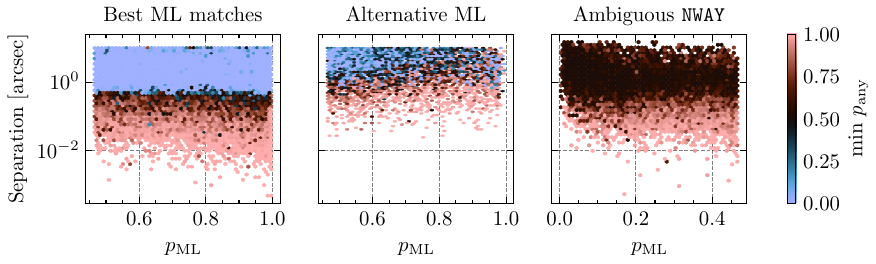}
    \caption{Comparing ML cross-matches with separation-based \nway{}.  Each panel shows a hexbin plot 
    of the cross-match separations as a function of $\pML$, with the color of the point in each bin represented by $\pANY$ (see color scale bars at right); the upper row of panels depict the maximum of $\pANY$, the middle row the median, and the bottom row the minimum.  The left column panels represent the $\approx113$k best ML cross-matches, the middle column that of $\approx8$k cases with multiple \gaia\ cross-matches, and the right column the $\approx22$k \nway{} cross-matches without good ML matches.
    }
    \label{fig:separation_vs_p_ML}
\end{figure*}

The contents of the catalog are displayed for various characteristics in Figures~\ref{fig:hr_diagram_best_matches_q}, \ref{fig:hr_diagram_test_pos_hardness}, and \ref{fig:separation_vs_p_ML}.  The first shows a color-magnitude HR diagram, the second shows the X-ray hardness ratios, and the last shows a comparison with \nway{} cross-matching.  They are each made using a hexagonal binning scheme to construct densities\footnote{Using the default binning scheme implemented in {\tt matplotlib.pyplot.hexbin}.  The advantage of hexagonal bins over regular rectangular bins is that hexagons allow for a better sampling efficiency for irregular shapes.}, and where applicable, we show plots made for both the full range (the minimum and maximum of the samples in each bin) as well as for a representative value (the median).

Figure \ref{fig:hr_diagram_best_matches_q} shows the Hertzsprung-Russell (HR) diagram for the best ML matches, the alternative ML, and the ambiguous \nway{} catalogs. In the Best and Alternative ML matches, the classical sequence shapes appear, and the distinct white‐dwarf cluster appears at blue colors and faint magnitudes.  In the \nway{} only matches, the HR diagram is washed out by spurious or poorly constrained matches, although the main sequence stars branch can be seen.  

Figure \ref{fig:hr_diagram_test_pos_hardness} again shows the HR diagram, but color coded by the X-ray spectral hardness $HR_{HS}$ for the CSC passbands $H=2-8$~keV and $S=0.5-1.2$~keV.  Redder colors represent softer spectra and bluer ones harder spectra.  Because each bin is comprised of several sources, we represent the spread in the hardness ratios with the median (middle row of panels) and the range (the minima in the bottom row, and the maxima in the top row); the columns represent the different types of matches (for Tables~\ref{tab:best_matches}, \ref{tab:alternative_matches}, and \ref{tab:ambiguous_matches} from left to right respectively).  The stellar main sequence is clearly visible in the median $HR_{HS}$ of best matches, and there is no structure visible for the \nway{}-only matches.  The difference in morphology going from the population of best ML matches to no ML matches is consistent with the assumption that underlies this work, that optical properties are indeed informative of X-ray emission. 

Figure~\ref{fig:separation_vs_p_ML} shows a direct comparison of the cross-matching between \nway{} and ML.  The panels are arranged in the same manner as in Figure~\ref{fig:hr_diagram_test_pos_hardness}, but depicts the variation in separation as a function of the ML score $\pML$, color coded by $\pANY$.  Notice that the x-axis range is $0.466\le\pML\le{1}$ for the left and middle columns which show ML identified matches, and is $0\le\pML<0.466$ for the right column since these matches do not contain any ML identified matches\footnote{Note that Table~\ref{tab:ambiguous_matches} includes some instances where \texttt{flag\_ml\_confident}=True \& \texttt{flag\_sep\_ok}=0.  These are not counted as valid ML matches.}.  As expected, matches found at larger separations (always for large off-axis locations $\theta$, per Equation~\ref{eqn:rmax}) tend to have smaller values of $\pANY$, but are uncorrelated with $\pML$.  The separations of the alternate matches are skewed to higher values (mean separation of $3''.5$ and a mean $\theta$ of $9'$).  The individual distributions of $\pML$ are all heavily skewed, with $50$\% of the best ML matches at $\pML>0.89$, of the alternate matches at $\pML>0.62$, and of the \nway{} only matches at $\pML<0.25$.  

A large fraction of the ML identified cross-matches are stars or galactic X-ray binaries.  Over $68$\% of the \gaia\ sources that correspond to the best ML match have valid parallaxes, and $41$\% of these are within $1$~kpc, with $94$\% within $8$~kpc.  The median distance to sources with valid parallaxes is $1.3$~kpc, suggesting that \chandra\ is seeing deep into the Galaxy.  In contrast, among the \nway{}-only no-ML cross-matches, $59$\% have valid parallaxes, and only $21$\% are within $1$~kpc, with a median distance of $2.1$~kpc, suggesting that even if these are real counterparts, they form a distinctly different population. 

\section{Discussion}\label{sec:discuss}

\subsection{Interpreting $\pML$}\label{sec:interpret}

Our model computes a quantity $\pML \in [0,1]$ that is a measure of whether a \gaia\ source near an X-ray source is a plausible counterpart to it.  This measure should not be interpreted as a probability, but as a score.  Higher numbers indicate that the likelihood is higher that the objects are indeed counterparts.

Note that $\pML$ is dependent only on the observable properties of the sources and not on their separation, unlike \nway{}.  Thus, it is often the case that when there are multiple optical candidates available for a given X-ray source, the match candidate at a larger separation may have a {higher} %larger 
ML score than the one that is situated closer.  We interpret this as an indication that the closer counterpart is a chance coincidence, and that the one with the higher $\pML$ is more likely to be the correct match, since it has properties that are more in line with the positive set in the training data.  We thus list all of the best ML candidates in one table (Table~\ref{tab:best_matches}).

It is interesting to consider the nature of the ambiguous matches. There are $7\,207$ alternative ML associations in which a \gaia{} candidate has a valid ML score but is not the highest-scoring ML counterpart for that CSC source (Table~\ref{tab:alternative_matches}).  When the ML score suggests a match with an X-ray source, but the identity of the X-ray source is unclear, we interpret the ML score as suggesting a potential X-ray source which may be detected with longer exposures or would have been detected if observed at a different epoch.  That is, we {\sl predict} X-ray emission from these \gaia\ sources.  {In addition, there are $\gtrsim{200}$ \gaia{} sources that are selected as the best ML counterpart for more than one CSC source.  These may occur due a combination of chance associations in dense fields, proper motion effects, or incompleteness in \gaia.}

The third catalog table (Table~\ref{tab:ambiguous_matches}) lists CSC sources that do not have ML identified counterparts, but do have proximate \gaia\ sources that are identified by \nway{} as matches.  A large fraction of them are likely to be chance coincidences: $\approx{50}$\% of these sources are at galactic latitudes $|b|<2^o$, compared to $\approx{25}$\% among the best ML matches.  Conversely, a similar number are very close coincidences, with $8\,595$ having $\pANY>0.9$ and $\textrm{separation}<0.5''$, and Ockham's razor suggests that these must be true counterparts, even though ML assigns a low score to them.  We interpret this as a systematic completeness error because our method relies primarily on \gaia, and sources that may be outliers in \gaia\ would be trained out of having a high ML score.  Thus, we expect that $\approx{7}$\% of the ML matches (or non-matches) could be in error.
{If these 8.6k sources, matched by \nway{} but not by ML, are illustrative of the systematic error in the 113k sources matched by ML, then we expect the error rate to be $\approx{7}$\%.}

\subsection{A Framework}\label{sec:framework}

While we have developed this cross-matching framework for the specific case of CSC and GDR3, our process is generalizable to cross-matching other catalogs. We describe here a framework that can be used for any arbitrary set of catalogs, which can be broken down into five steps:

\paragraph{Perform a primary spatial cross-match} Bring together a primary catalog (e.g., CSC2.1) and a candidate counterpart catalog (e.g., \gaia\ DR3), then run a spatial match, keeping every object within some radius. This provides a list of potential counterparts, as couples, for the primary catalog.

\paragraph{Create a positive and negative set} Define a handful of high‐confidence true matches (small separation, high positional-overlap probability) and obvious non‐matches (large separation, low probability or random picks). Those become the positive and negative examples for the next step.

\paragraph{Train a probabilistic classifier} Train a machine‐learning classifier (e.g., gradient-boosted trees) on all the useful multi-wavelength and catalog properties. In our specific implementation, this includes fluxes, colors, proper motions, variability flags, even external class labels. Run that model over every candidate, producing a “match score” between 0 and 1.

\paragraph{Define your thresholds} Define a threshold for a reliable counterpart taking into account the information available: spatial probabilities (\nway{}), machine learning scores, separation, and the primary source off-axis. With this, define a set of quality labels.

\paragraph{Make a final selection} Collapse all possible counterparts back into one reliable counterpart per primary source by taking, for each primary source detection, the best candidate source within the thresholds defined before.

Despite this work being heavily oriented towards \chandra\ and \gaia, we bring here some ideas that might be applicable for any cross-matching between two catalogs. 
The most challenging step is selecting appropriate data for training a classifier {(i.e., constructing positive and negative sets without the benefit of labeled ground truth data).  We suggest using statisics-based heuristsics,}
as shown in Section~\ref{sec:limits_to_NWAY}, to select an appropriate cutoff for the construction of positive and negative sets.

\section{Summary}\label{sec:summary}
\begin{figure*}
    \centering
    \includegraphics[width=\linewidth]{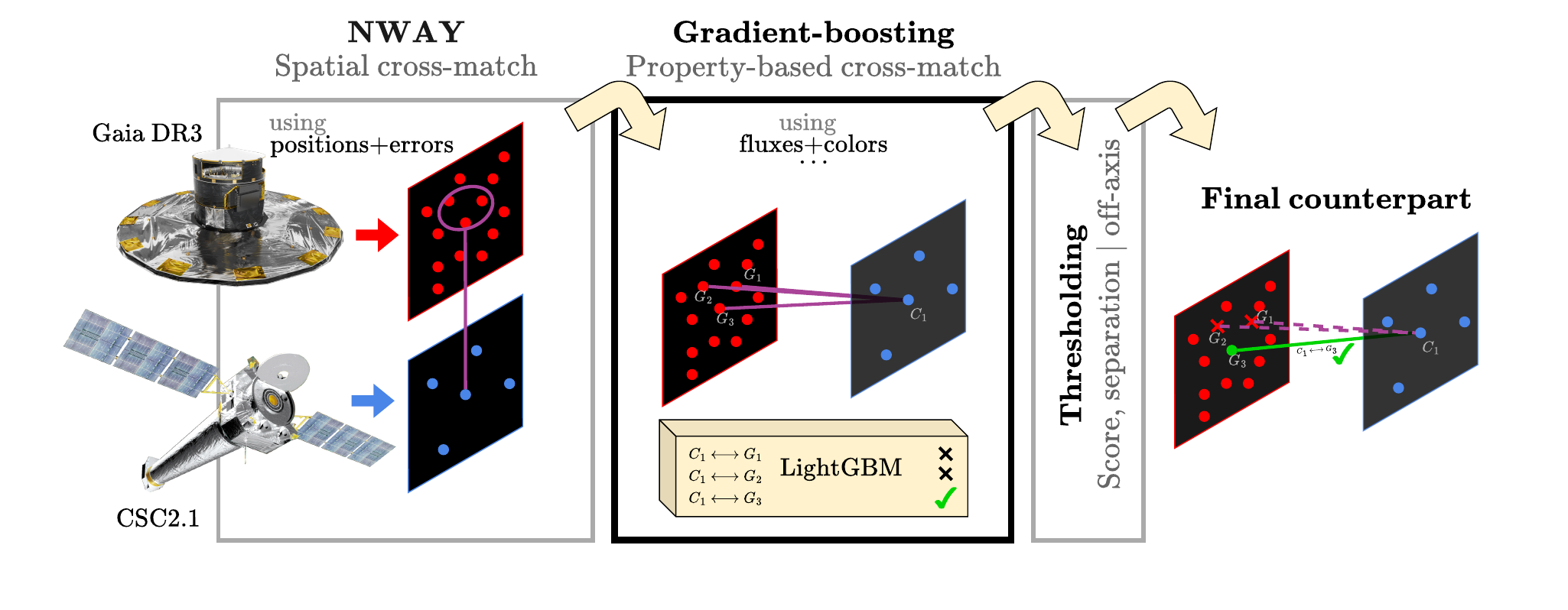}
\caption{%
    {Cross-matching the \chandra\ Source Catalog 2.1 and \gaia\ DR3.}  
    \textcolor[HTML]{FF0000}{Red} and \textcolor[HTML]{4A85E8}{blue} dots represent \textcolor[HTML]{FF0000}{\gaia} and \textcolor[HTML]{4A85E8}{\chandra} sources, in the upper and lower projected squares, respectively. 
    In the \textsl{first stage}, a spatial crossmatch is performed using \nway{}, providing a first set of \textcolor[HTML]{AA4499}{candidate associations}. 
    In the \textsl{second stage}, each candidate pair is scored using a \lgbm{} classifier trained on catalog features from \textcolor[HTML]{4A85E8}{\chandra} and \textcolor[HTML]{FF0000}{\gaia}. 
    \textsl{Finally}, candidates are filtered using score and separation thresholds to select the most likely astrophysical counterpart. 
    The \textcolor[HTML]{00D402}{\textsl{green checkmark}} indicates the final accepted match. 
    Doing this for all potential associations allows us to construct a catalog of counterparts.
}
\label{fig:schema}
\end{figure*}

We have developed a machine learning-based framework to cross-match X-ray with optical sources.  The defining feature of our method is that it primarily leverages properties of the sources like magnitudes and colors to decide whether they are likely to be matches.  Our method defines a general process (illustrated in Figure~\ref{fig:schema}) to transcend source separation as the sole criterion for cross-matching, and helps to resolve ambiguities and flag chance coincidences.  We apply it to cross-match X-ray sources in the \chandra\ Source Catalog with optical sources from \gaia~DR3.

We base our method on first defining a positive set of ${\approx}30$k cross-matches using our best knowledge of X-ray and optical matches, which relies primarily on separation and is validated by observable differences in optical magnitudes (Section~\ref{sec:pos_set}).  We construct a negative set of size ${\approx}310$k using unambiguously determinable non-matches (Section~\ref{sec:neg_set}).  We then train the model using a large variety of features extracted from both catalogs (see Table~\ref{tab:features}), using a framework that relies {on gradient boosting with} {\tt LightGBM} (see Section~\ref{sec:implementation}).  However, the general approach is model-agnostic, and other architectures, including deep learning, could be explored for the cross-matching task.  We determine a threshold score for a match using both astronomical (discarding plausible chance coincidences) and empirical (using ROC-AUC curves) arguments, which happen to be consistent with each other: ML scores $\pML>0.466$ are accepted as valid cross-matches (Section~\ref{sec:counterpart_selection}).

We test the method on the well-studied \chandra\ dataset of the Orion Nebula Cluster (Section~\ref{sec:verification}), and find that the bare ML method matches \nway{} cross-matching results at a high level of accuracy (see Table~\ref{tab:nway_ml_coup}).  Indeed, 95\% of \nway{} cross-matches are reproduced by ML matches, which is a remarkable result, since the former strictly uses separation and the latter uses similarity in features.  Using this method, we then generate a new catalog of cross-matches between serendipitously detected \chandra\ X-ray sources from \chandracat{} and optical sources from \gdrversion{} (see Section~\ref{sec:catalog}).  We present the cross-matches in three tables: the first listing the ${\approx}113$k cross-matches with the best ML score ($\mathrm{max }$ $\pML$; Table~\ref{tab:best_matches}); the second listing the ${\approx}8000$ plausible alternative ML cross-matches ($0.466\le\pML<\mathrm{max}\text{ }\pML$; Table~\ref{tab:alternative_matches}); and the third listing the ${\approx}22$k cross-matches which have such small separations that \nway{} registers a high probability of a match,
{even though they fail the ML match criterion ($\pANY>0.5$ and $\pML<0.466$; Table~\ref{tab:ambiguous_matches}).  The small separations suggest that many of them could be real matches, and a definitive decision requires a more detailed analysis (see also discussion of systematic errors in Section~\ref{sec:interpret} above).}
The full catalog enables future detailed study of subsets of object classes, like stars, which are abundantly represented in \gaia.

Our method can be generalized into a practical framework that can be applied to other catalog combinations (Section~\ref{sec:discuss}).  Our goal here is not to replace established spatial-separation dependent methods like \nway{}, but to supplement them by incorporating other properties of the catalog objects.  In this, we improve upon methods that incorporate the distributions of such properties as priors.  Our method enables the recovery of correct matches even in unusual scenarios, such as contamination from chance interlopers, or when there are systematic errors present in the measured coordinates (e.g., due to influence from nearby sources, or PSF distortions at large off-axis angles, or uncorrected proper motion).

\begin{acknowledgments}
{\sl Acknowledgements:} We thank Joshua Wing, Raffaele D'Abrusco, Martina Cadiz, Daniel Moreno-Cartagena, and Kevin Ortiz-Ceballos for useful discussions and comments throughout this research. We are grateful for the support of AstroAI (\url{https://astroai.cfa.harvard.edu}), a new center developing artificial intelligence methods to enable next generation astrophysics at the CfA. {This work is supported by the National Science Foundation under Cooperative Agreement PHY-2019786 (The NSF AI Institute for Artificial Intelligence and Fundamental Interactions, http://iaifi.org/).} We acknowledge support from \chandra\ grant AR3-24002X and from the NASA Contract to the \chandra\ X-Ray Center NAS8-03060. This research has made use of data obtained from the \chandra\ Source Catalog, provided by the \chandra\ X-ray Center (CXC). This work has made use of data from the European Space Agency (ESA) mission \gaia\ (\url{https://www.cosmos.esa.int/gaia}), processed by the {\gaia} Data Processing and Analysis Consortium (DPAC, \url{https://www.cosmos.esa.int/web/gaia/dpac/consortium}). Funding for the DPAC has been provided by national institutions, in particular the institutions participating in the {\gaia} Multilateral Agreement.
This material is based upon work supported by the National Science Foundation Research Fellowship Program under Grant No DGE2140739. Any opinions, findings, and conclusions or recommendations expressed in this material are those of the authors and do not necessarily reflect the views of the National Science Foundation.
\end{acknowledgments}

\facilities{CXO, Gaia}

\software{LightGBM \citep{kelightgbmhighlyefficient2017}, scikit-learn \citep{pedregosascikitlearnmachinelearning2011}, Astropy \citep{collaborationastropyprojectsustaining2022}, PINTofALE \citep{2000BASI...28..475K}.}

\bibliography{references}{}
\bibliographystyle{aasjournal}

\appendix

\section{Brief description of \nway{}}
\label{app:nway}

\nway{} is a Bayesian algorithm designed to cross-match astronomical sources across multiple catalogs simultaneously. Building on the formalization introduced by \cite{budavariprobabilisticcrossidentificationastronomical2008}, it combines astrometric information (source positions and uncertainties) and photometric priors (e.g., magnitude or color distributions) to calculate the probability that each candidate source is the correct counterpart. For each candidate counterpart \( i \) of a given primary source, \nway{} computes the posterior probability \nwaypi{} that candidate \( i \) is the correct match, defined as:

\[
\nwaypi = \frac{P(H_i|D)}{\sum_{j} P(H_j|D)},
\]

where \( P(H_i|D) \) is the posterior probability of hypothesis \( H_i \) (i.e., candidate \( i \) is the true counterpart) given the observed data \( D \), which incorporates positional and photometric information. Additionally, \nway{} calculates \nwaypany{}, the probability that any of the candidate sources is a true counterpart:

\[
\nwaypany{} = 1 - \frac{P(H_0|D)}{\sum_{j} P(H_j|D)},
\]

where \( P(H_0|D) \) is the posterior probability for the hypothesis \( H_0 \), meaning none of the candidates correspond to the primary source. For details on the formalism, we refer the reader to \cite{salvatofindingcounterpartsallsky2018, budavariprobabilisticcrossidentificationastronomical2008}.

\section{Model Training Details}
\label{app:model_training}

\subsection{Model Configuration}

\paragraph{Preprocessing}  
Feature preprocessing was intentionally lightweight. For numerical features, we apply a $\log_{10}(1 + x)$ transformation only to variables with highly skewed distributions or large dynamic ranges. Specifically, the following features are log-transformed: 

\begin{itemize}
  \item \texttt{parallax} %($\plx$)
  \item \texttt{parallax\_error}
  \item \texttt{photflux\_aper\_b}
  \item \texttt{phot\_g\_mean\_flux}
  \item \texttt{phot\_bp\_mean\_flux}
  \item \texttt{phot\_rp\_mean\_flux}
  \item \texttt{radial\_velocity}
  \item \texttt{distance\_gspphot}
  \item \texttt{sqrt(pmra\string^2 + pmdec\string^2)}
\end{itemize}

These transformations are applied to all rows, using the expression $\log_{10}(1 + x)$ to avoid undefined values for zero. Categorical features like \texttt{yangetal\_gcs\_class}, \texttt{yangetal\_training\_class}, \texttt{perezdiazetal\_class}, and \texttt{phot\_variable\_flag} are passed to LightGBM as raw \texttt{category}-type columns. Missing data is not imputed and kept as \texttt{nan}. Missing Gaia and spatial properties in the random negatives are filled with the value \texttt{-1}.

\paragraph{Hyperparameter Tuning}  
We perform hyperparameter tuning using \texttt{RandomizedSearchCV} with 200 iterations over a custom-defined search space, using 5-fold cross-validation and AUC-ROC as the scoring metric. Each parameter is sampled either log-uniformly or uniformly, and integer constraints are enforced where appropriate. The full hyperparameter space included:

\begin{itemize}
  \item \texttt{learning\_rate}: log-uniform over $[e^{-7}, 1]$
  \item \texttt{num\_leaves}: quantized log-uniform over $[e^0, e^7]$, step size 1, integer
  \item \texttt{feature\_fraction}: uniform over $[0.5, 1.0]$
  \item \texttt{bagging\_fraction}: uniform over $[0.5, 1.0]$
  \item \texttt{min\_data\_in\_leaf}: quantized log-uniform over $[e^0, e^6]$, step size 1, integer
  \item \texttt{min\_sum\_hessian\_in\_leaf}: log-uniform over $[e^{-16}, e^{5}]$
  \item \texttt{lambda\_l1}: random choice from \{0, log-uniform in $[e^{-16}, e^2]$\}
  \item \texttt{lambda\_l2}: random choice from \{0, log-uniform in $[e^{-16}, e^2]$\}
\end{itemize}

For the tuning run, we set the number of boosting rounds to 500. The best model configuration is selected after evaluating all 200 trials. Afterward, the number of boosting rounds is fixed at $5000$ and early stopping was set with a patience of 10 rounds. The final model is retrained using these parameters and evaluated on the held-out test set.

We also treat data construction choices as part of the hyperparameter search. In particular, we run separate tuning experiments to assess the effect of including intermediate negatives and to determine the optimal number of random negatives to sample per unique \chandra\ source. For the latter, we explore multipliers of {0, 5, 10, 20, 40, 60, 80, 100} relative to the number of CSC sources. For each configuration, we perform a full hyperparameter tuning run as described above. We find that including intermediate negatives consistently improved model performance. As for the random negatives, we find the best performance when using a multiplier of 5, which is ultimately adopted in the final configuration.

We experiment with different methods of hyperparameter search, and also different configurations of the number of search-iterations and folds in the cross-validation. For simplicity, we report here the one that we ultimately selected and chose a model from.

\paragraph{Final Model}  
The best-performing model was trained with the following parameters (rounded):

\begin{itemize}
  \item \texttt{learning\_rate = 0.00456}
  \item \texttt{num\_leaves = 131}
  \item \texttt{n\_estimators = 5000}
  \item \texttt{min\_data\_in\_leaf = 2}
  \item \texttt{min\_sum\_hessian\_in\_leaf = 5.66$\tt \times 10^{-7}$}
  \item \texttt{bagging\_fraction = 0.914}
  \item \texttt{feature\_fraction = 0.519}
  \item \texttt{lambda\_l1 = 0}
  \item \texttt{lambda\_l2 = 0.205}
\end{itemize}

\paragraph{Performance} The model achieved an AUC of $0.894$ on the validation set and $0.898$ on the independent test set. The model was later saved and used for the analysis presented in the paper.

\section{Thresholds from \chandra's PSF}
\label{app:thresholds}
First, the astrometric precision of \chandra\ must be accounted for. Typical uncertainties in source positions are about 0.5\arcsec, comparable to the ACIS pixel size, although they can be reduced to 0.1--0.3\arcsec\ in fields with deep observations and available re-registrations.

Second, the statistical error in locating a source depends on photon counts. The  positional uncertainty is expected to scale as PSF width divided by $\sqrt{N}$. For faint sources ($\sim$10 counts), we expect a rounded statistical error of at least $\sim\text{PSF}/3$.

Third, proper motion effects become significant over decade-long baselines between \chandra\ and \gaia\ observations. {The median proper motion of all the \gaia\ potential cross-match candidates is $\approx$5~mas~yr$^{-1}$, which can accumulate to $\sim$0.1\arcsec, adding another term to the positional uncertainty.} 
%Typical differences reach $\sim$0.1\arcsec, adding another term to the uncertainty.

Fourth, unresolved overlapping sources can dominate in crowded fields. If the overlap is less than the PSF width, the sources might be impossible to resolve with any method. If variability is present, the sources could show fundamental differences between epochs. This adds an additional uncertainty roughly at the PSF scale.

Section \ref{sec:limits_to_NWAY}  sets the scale for reasonable maximum separations near the \chandra\ aimpoint. However, the \chandra\ PSF broadens significantly as a function of the off-axis angle $\theta$. Empirically, we adopt a parabolic parametrization of the encircled energy radii as a function of off-axis, from the Figure 4.12 of \chandra\ Proposer's Observatory Guide,\footnote{\url{https://cxc.harvard.edu/proposer/POG/html/chap4.html\#fg:hrma\_ee\_offaxis\_hrci}}
\begin{equation}
R^{\text{ECF}}_{90}(\theta) = 1.1\arcsec + 0.05\arcsec \times \theta + 0.1\arcsec \times \theta^2 ~\,~\mathrm[arcsec]
\label{eqn:R_ECF}
\end{equation}
where $R^{\text{ECF}}_{90}$ is the radius (in arceconds) containing 90\% of the PSF's encircled energy at off-axis angle $\theta$ (in arcminutes). For a circular Gaussian PSF of width $\sigma$, the $90\%$ encircled energy radius corresponds to around $2.15\sigma$. Thus, using $\sigma_{\text{PSF}} \approx R^{\text{ECF}}_{90}/2.15$, we estimate the PSF contribution at each off-axis location. 
\section{Columns of counterparts catalog}
\label{app:columns}
Each row in the catalog tables includes the following columns:

\begin{itemize}
  \item \texttt{csc21\_name} – Unique identifier of the \chandra\ source.
  \item \texttt{csc21\_ra}, \texttt{csc21\_dec} – Right ascension and declination of the \chandra\ source (in degrees).
  \item \texttt{min\_theta\_mean} – Mean off-axis angle of the \chandra\ detection (in arcminutes).
  \item \texttt{gaia3\_source\_id} – Unique identifier of the \gaia\ DR3 source.
  \item \texttt{gaia3\_ra}, \texttt{gaia3\_dec} – Right ascension and declination of the \gaia\ source (in degrees).
  \item \nwaypi{} – \nway{} positional match probability for the CSC-\gaia\ pair.
  \item \nwaypany{} – \nway{} probability that the \chandra\ source has any counterpart.
  \item \pmatchindvar{} – $\pML$, Machine-learning score estimating the likelihood of a true match.
  \item \texttt{separation} – Angular separation between the CSC and \gaia\ positions (in arcseconds).
  \item \texttt{match\_flag} – Original label from \nway{}, used in training/validation.
  \item \texttt{flag\_nway\_confident} – True if $\pANY \geq$ 0.5 for the \gaia\ source with highest \nwaypi{}.
  \item \texttt{flag\_ml\_confident} – True if $\pML \geq$ 0.466.
  \item \texttt{flag\_sep\_ok} – True if the pair satisfies the separation and off-axis angle thresholds.
  \item \texttt{phot\_g\_mean\_mag} - \gaia\ counterpart's G-band apparent magnitude.
\end{itemize}

\clearpage
\section{Catalog Samples}\label{app:catalog_samples}

We show representative samples of the catalog tables (see Section~\ref{sec:catalog}) as for the best ML matches (Table~\ref{tab:best_matches}), but for alternative ML matches (Table~\ref{tab:alternative_matches}), and for \nway{}-only matches that do not have reliable ML counterparts (Table~\ref{tab:ambiguous_matches}).  The full tables are available on \citep{perez_diaz_2026_18652667}.

\begin{deluxetable*}{ccccccccccccc}
\tablecaption{\textit{Alternative ML Counterpart Matches:}\\
As Table~\ref{tab:best_matches}, for an illustrative sample of counterparts from the Alternative ML matches catalog. These 10 entries (from 7\,242 total) have $\pML{} \geq 0.466$ and separation criteria met, but do not have the highest ML score (i.e., $\pML\ne\max \pML{}$) for any CSC counterpart.}\label{tab:alternative_matches}
\tablewidth{0pt}
\tablehead{
\colhead{Name} & \colhead{RA} & \colhead{Dec} & \colhead{$\theta_0$} & \colhead{Gaia ID} & \colhead{$\pNWAY{}$} & \colhead{$\pANY{}$} & \colhead{$\pML{}$} & \colhead{Sep.} & \colhead{$F_N$} & \colhead{$F_M$} & \colhead{$F_S$} & \colhead{$\gmag$} \\
 & \colhead{[deg]} & \colhead{[deg]} & \colhead{[arcmin]} &  &  &  &  & \colhead{[arcsec]} &  &  &  & \colhead{[mag]}
}
\startdata
000102.5+672840 & 0.26 & 67.48 & 8.4 & 528573584943861632 & $6.6 \times 10^{-13}$ & 0.99 & 0.49 & 3.7 & \xmark & \cmark & \cmark & 19.3 \\
043340.8+174440 & 68.42 & 17.74 & 7.5 & 3313485355247771264 & 0.49 & 0.98 & 0.79 & 1.2 & \xmark & \cmark & \cmark & 11.8 \\
054044.2-690245 & 85.18 & -69.05 & 11.5 & 4657683743840276352 & $5.3 \times 10^{-4}$ & 0.63 & 0.50 & 6.8 & \xmark & \cmark & \cmark & 19.8 \\
103711.4-583443 & 159.30 & -58.58 & 7.3 & 5350683352738606720 & $2.9 \times 10^{-19}$ & 0.92 & 0.61 & 3.9 & \xmark & \cmark & \cmark & 17.5 \\
131521.8-552259 & 198.84 & -55.38 & 13.9 & 6066625779323513984 & 0.20 & 0.60 & 0.61 & 4.6 & \xmark & \cmark & \cmark & 11.8 \\
163921.1-474551 & 249.84 & -47.76 & 5.1 & 5941070966009730816 & 0.28 & 0.83 & 0.50 & 2.4 & \xmark & \cmark & \cmark & 18.3 \\
173947.1-394342 & 264.95 & -39.73 & 8.3 & 5960785548794278912 & $5.8 \times 10^{-6}$ & 0.66 & 0.66 & 4.0 & \xmark & \cmark & \cmark & 19.7 \\
175831.5-290116 & 269.63 & -29.02 & 6.5 & 4062352142152258688 & $3.3 \times 10^{-4}$ & 0.48 & 0.49 & 3.4 & \xmark & \cmark & \cmark & 17.1 \\
190319.3+030027 & 285.83 & 3.01 & 12.1 & 4268770370618299264 & 0.11 & 0.46 & 0.48 & 7.2 & \xmark & \cmark & \cmark & 19.5 \\
235802.7+230354 & 359.51 & 23.07 & 8.3 & 2848261339629138176 & 0.35 & $>$0.999 & 0.87 & 0.68 & \xmark & \cmark & \cmark & 14.9 \\
\enddata
\end{deluxetable*}

\begin{deluxetable*}{ccccccccccccc}
\tablecaption{\textit{Ambiguous \nway{} Matches:}\\
As Table~\ref{tab:best_matches}, for an illustrative sample of counterparts from the Ambiguous \nway{} matches catalog. These 10 entries (from 19\,649 total) have $\pANY{} \geq 0.5$ indicating highly probable association based on separation, but do not have ML matches.}\label{tab:ambiguous_matches}
\tablewidth{0pt}
\tablehead{
\colhead{Name} & \colhead{RA} & \colhead{Dec} & \colhead{$\theta_0$} & \colhead{Gaia ID} & \colhead{$\pNWAY{}$} & \colhead{$\pANY{}$} & \colhead{$\pML{}$} & \colhead{Sep.} & \colhead{$F_N$} & \colhead{$F_M$} & \colhead{$F_S$} & \colhead{$\gmag$} \\
 & \colhead{[deg]} & \colhead{[deg]} & \colhead{[arcmin]} &  &  &  &  & \colhead{[arcsec]} &  &  &  & \colhead{[mag]}
}
\startdata
000001.5-245151 & 0.01 & -24.86 & 17.1 & 2335109264262060928 & $>$0.999 & 0.93 & 0.10 & 5.3 & \cmark & \xmark & \cmark & 17.9 \\
052301.2+332411 & 80.76 & 33.40 & 4.7 & 180986146577295616 & $>$0.999 & 0.997 & 0.29 & 0.17 & \cmark & \xmark & \cmark & 20.1 \\
102443.6-573714 & 156.18 & -57.62 & 10.6 & 5351757438154160000 & $>$0.999 & 0.81 & 0.38 & 1.7 & \cmark & \xmark & \cmark & 14.3 \\
111048.2-372645 & 167.70 & -37.45 & 9.7 & 5397353940526126080 & 0.999 & 0.87 & 0.42 & 1.7 & \cmark & \xmark & \cmark & 20.7 \\
141217.2-513259 & 213.07 & -51.55 & 7.5 & 6089551348357034752 & $>$0.999 & 0.65 & 0.04 & 1.2 & \cmark & \xmark & \cmark & 19.4 \\
164824.8-461016 & 252.10 & -46.17 & 3.8 & 5940085902338251264 & $>$0.999 & 0.98 & 0.40 & 0.02 & \cmark & \xmark & \cmark & 17.4 \\
174505.2-294511 & 266.27 & -29.75 & 8.0 & 4056882454043395712 & 0.54 & 0.70 & 0.12 & 0.35 & \cmark & \xmark & \cmark & 18.1 \\
180501.5-273917 & 271.26 & -27.65 & 2.4 & 4062934780448760064 & $>$0.999 & 0.87 & 0.26 & 0.11 & \cmark & \xmark & \cmark & 18.2 \\
191630.5-155553 & 289.13 & -15.93 & 9.4 & 4184142468158354048 & 0.96 & 0.67 & 0.42 & 1.7 & \cmark & \xmark & \cmark & 19.1 \\
235928.0+440317 & 359.87 & 44.05 & 9.2 & 1922753577342779904 & $>$0.999 & 0.97 & 0.35 & 1.0 & \cmark & \xmark & \cmark & 20.7 \\
\enddata
\end{deluxetable*}

\end{document}